\documentclass[spanningrule,conferencestyle]{cambridge6B}
\usepackage{natbib}
\usepackage[figuresright]{rotating}
\usepackage{floatpag}
\rotfloatpagestyle{empty}
\usepackage{amsthm}
\usepackage{graphicx}
\usepackage{makeidx}
\makeindex
\theoremstyle{plain}
\newtheorem{theorem}{Theorem}[section]

\newtheorem{proposition}[theorem]{Proposition}

\newtheorem*{theorem*}{Theorem}
\newtheorem*{lemma*}{Lemma}
\newtheorem*{proposition*}{Proposition}
\newtheorem*{corollary*}{Corollary}
\newtheorem*{conjecture*}{Conjecture}
\theoremstyle{definition}
\newtheorem{definition}[theorem]{Definition}
\newtheorem{example}[theorem]{Example}

\newtheorem*{definition*}{Definition}
\newtheorem*{example*}{Example}
\newtheorem*{prob*}{Problem}
\newtheorem*{remark*}{Remark}
\newtheorem*{notation*}{Notation}
\newtheorem*{exer*}{Exercise}
\def\makeRRlabeldot#1{\hss\llap{#1}}
\renewcommand\theenumi{{\rm (\roman{enumi})}}
\renewcommand\theenumii{{\rm (\alph{enumii})}}
\renewcommand\theenumiii{{\rm (\arabic{enumiii})}}
\renewcommand\theenumiv{{\rm (\Alph{enumiv})}}

\usepackage{amsfonts}
\usepackage{nicefrac}

\usepackage{pgfplots}  

\usepackage{amsmath}
\usepackage{amssymb}
\usepackage{bm}
\usepackage{enumitem}
\usepackage{listings}
\usepackage[section]{placeins}
\usepackage{xcolor}
\usepackage{soul}
\usepackage{hyperref}

\usepackage{mathpartir}
\usepackage{myabbrevmacros}
\usepackage{mylangmacros}
\usepackage{myprobmacros}
\usepackage{mysemmacros}
\usepackage{myutilmacros}

\newcommand{\mathcolorbox}[2]{\colorbox{#1}{$\displaystyle #2$}}

\colorlet{shadecolor}{gray!40}
\newcommand{\mlhighlight}[1]{\mathcolorbox{shadecolor}{#1}}

\newcommand{\two}[0]{\ensuremath{\bm{2}}}
\newcommand\sampfunc{\ensuremath{\bm{S}}}
\newcommand\probfunc{\ensuremath{\bm{P}}}

\begin{document}


\mainmatter
\setcounter{page}{1}

\author[Huang, Morrisett and Spitters]
{Daniel Huang\\
University of California, Berkeley \and
Greg Morrisett\\
Cornell University \and
Bas Spitters\\
Aarhus University}

\chapter[An Application of Computable Distributions \dots]{An Application of Computable Distributions to the Semantics of Probabilistic Programs}

\vspace{0.5cm}
\chapterabstract 
In this chapter, we explore how (Type-2) computable distributions can be used to give both (algorithmic) sampling and distributional semantics to probabilistic programs with continuous distributions. Towards this end, we sketch an encoding of computable distributions in a fragment of Haskell and show how topological domains can be used to model the resulting PCF-like language. We also examine the implications that a (Type-2) computable semantics has for implementing conditioning. We hope to draw out the connection between an approach based on (Type-2) computability and ordinary programming throughout the chapter as well as highlight the relation with constructive mathematics (via realizability).

\chapterkeywords{computable distributions, semantics, probabilistic programs, topological domains, realizability}


\definecolor{fadednavy}{HTML}{444CFC}
\definecolor{punch}{HTML}{CE5A57}
\definecolor{oceanbreeze}{HTML}{78A5A3}
\definecolor{warm}{HTML}{E1B16A}


\lstnewenvironment{hlisting}
  { \lstset{
      language=Haskell,
      basicstyle=\small\ttfamily,
      breaklines=true,
      xleftmargin=.05\textwidth,
      keywords=[1]{let, where, do, return, in, class, instance, type, newtype, if, then, else, case, of},
      keywords=[2]{forall, Real, Samp, Rat, Integer, CMetrizable, A, BndDens, MList, Either, Nat, IO, Vec},
      keywords=[3]{import, module},
      keywordstyle={[1]\bfseries\ttfamily\color{fadednavy}},
      keywordstyle={[2]\bfseries\ttfamily\color{punch}},
      keywordstyle={[3]\bfseries\ttfamily\color{fadednavy}},
      commentstyle=\color{oceanbreeze},
    }
    \
  }
  {
  }

\section{Overview}
\label{sec:over}

Probabilistic programs exhibit a tension between the \emph{continuous} and the \emph{discrete}. On one hand, we are interested in using probabilistic programs to model natural phenomena---phenomena that are often modeled well with \emph{reals} and \emph{continuous distributions} (\eg, as in physics and biology). On the other hand, we are also bound by the fundamentally \emph{discrete nature of computation}, which limits how we can (1) represent models as programs and then (2) compute the results of \emph{queries} on the model. The aim of this chapter\footnote{This chapter contains material from~\citet{huang2017pprog} and~\citet{huang2016sem}.}\footnote{Daniel Huang was supported by DARPA FA8750-17-2-0091.}\footnote{Bas Spitters was partially supported by the Guarded homotopy type theory project, funded by the Villum Foundation, project number 12386 and partially by the AFOSR project `Homotopy Type Theory and Probabilistic Computation', 12595060. Any opinions, findings and conclusions or recommendations expressed in this material are those of the authors and do not necessarily reflect the views of the AFOSR.} is to keep this tension in the fore by using the notion of a \emph{(Type-2) computable distribution} as a lens through which to understand probabilistic programs. We organize our exploration via a series of questions.
\begin{enumerate}[noitemsep]
    \item \emph{What is a (Type-2) computable distribution} (Section~\ref{sec:crev})? First, we review \emph{Type-2 computability}~\citep[\eg, see][]{weihrauch2000computable} and how it applies to reals and continuous distributions. The high-level idea is to represent continuum-sized objects as a sequence of discrete approximations that converge to the appropriate object instead of abstracting the representation of such an object.
    \item \emph{How do we implement continuous distributions as a library in a general-purpose programming language} (Section~\ref{sec:lib})? After we have seen the basic idea behind Type-2 computability, we sketch an implementation of reals and continuous distributions in a fragment of Haskell. We emphasize that the implementation does not assume any reals, continuous distributions, or operations on them as black-box primitives.
    \item \emph{What mathematical structures can we use to model such a library} (Section~\ref{sec:math})? Our next step is to find mathematical structures that can be used to faithfully model the implementation. Towards this end, we review \emph{topological domains}~\citep[\eg, see][]{battenfeld2007convenient,battenfeld2008td}, which are an alternative to traditional structures used in denotational semantics. Topological domains support all the standard domain-theoretic constructions needed to model PCF-like\footnote{Recall that the PCF (Programming Computable Functions) language is a core calculus that can be used to model typed functional languages such as Haskell.} languages as well as capture the notion of (Type-2) computability. In particular, we can encode reals and continuous distributions as topological domains so that they are suitable for our purposes of giving semantics.
    \item \emph{What does a semantics for a core language look like} (Section~\ref{sec:core})? In this section, we make the connection between the implementation and the mathematics more concrete by using the constructs described previously to give both (algorithmic) sampling and distributional semantics to a core PCF-like language extended with reals and continuous distributions (via a probability monad) called $\langprob$.\footnote{$\langprob$ also supports distributions on any countably based space. This means that $\langprob$ does not (in general) have distributions on function spaces, although the language itself contains higher-order functions.} The sampling semantics can be used to guide implementation while the distributional semantics can be used for equational reasoning.
    \item \emph{What are the implications of taking a (Type-2) computable viewpoint for Bayesian inference} (Section~\ref{sec:infer})? Perhaps surprisingly, at least to those who employ Bayesian inference in practice, it can be shown that \emph{conditioning} is not (Type-2) computable~\citep[see][]{ackerman2011noncomputable}. Hence, there is a sense in which a ``Turing-complete" probabilistic programming language cannot support conditional queries for every expressible probabilistic model. Fortunately, we do not run into these pathologies in practice and can recover conditioning in sufficiently general settings.
\end{enumerate}
We hope to draw out the connection between an approach based on (Type-2) computability with ordinary programming (\ie, programming in a fragment of Haskell) throughout the chapter as well as highlight the relation with constructive mathematics via \emph{realizability}~\citep[\eg, see][]{streicher2008realize} (Section~\ref{subsec:math:realize}).

\paragraph{Prerequisites}
We assume basic knowledge of programming language semantics~\citep[\eg, at the level of][]{gunter1992semantics}. For our purposes, this primarily includes (1) the application of category theory to programming language semantics and (2) the use of complete partial orders (CPOs) to model the semantics of PCF. As we will be giving examples in Haskell, familiarity with the Haskell programming language will also be assumed.\footnote{Familiarity with other typed functional languages such as ML should also suffice, although we should remind ourselves that Haskell has call-by-need semantics so that it has a lazy order-of-evaluation.} Finally, we assume basic knowledge of measure-theoretic probability~\citep[\eg, see][]{durrett2010}.

\section{Computability Revisited}
\label{sec:crev}

\emph{What is a computable distribution?} One approach to studying computability is based on Turing machines~\citep[\eg, see][]{sipser2012computation}. Under this approach, we define (1) a \emph{machine model} (\ie, the Turing machine) and (2) conditions under which the machine model is said to \emph{compute}. More concretely, a Turing machine is said to \emph{compute} a (partial) function $f: \Sigma^* \rightharpoonup \Sigma^*$ if it halts with $f(w) \in \Sigma^*$ on the output tape given $w \in \Sigma^*$ on an input tape, where $\Sigma$ is a finite set and $\Sigma^* \eqdef \set{a_0 \dots a_n \ST a_i \in \Sigma, 0 \leq i \leq n}$ is a collection of words comprised of characters from $\Sigma$. The two element set $\two \eqdef \set{0, 1}$ for bits (or booleans) is a commonly used alphabet.

This definition of computability reveals that traditional computation is fundamentally \emph{discrete}. We can see this directly in the definition of a computable function (with type $\Sigma^* \rightharpoonup \Sigma^*$), which maps elements of a discrete domain (\ie, a set of finite words $\Sigma^*$) to elements of a discrete codomain (\ie, a set of finite words $\Sigma^*$ again). As $\Sigma^*$ is countable, it cannot be put in bijection with the reals $\R$; hence, we cannot encode all the reals on a Turing machine.

One immediate issue that this highlights for probabilistic programs is how one should handle reals and continuous distributions while maintaining the connection back to computation. A pragmatic solution to this is to use floating point arithmetic, \ie, discretize and finitize the reals. From this perspective, we can model the semantics of probabilistic programs using floating point numbers and finitely-supported discrete distributions (on floats) so that the semantics more faithfully models an actual implementation. Nevertheless, we sacrifice the correspondence between the program and the mathematics that we use on pencil-paper. An alternative to the situation above is to generalize the notion of computability to continuum-sized sets in such a way that the computations can still by implemented by a standard machine.

\subsection{Type-2 Computability}
\label{subsec:crev:higher}

\emph{Type-Two Theory of Effectivity}~\citep[abbreviated TTE, see][]{weihrauch2000computable} changes the conditions under which a machine is said to compute an answer but keeps the machine model as is. In this setting, a machine is said to \emph{compute} a function $f: \Sigma^\omega \rightharpoonup \Sigma^\omega$ if it can write any initial segment of $f(w) \in \Sigma^\omega$ on the output tape in finite time given $w \in \Sigma^*$ on an input tape, where $\Sigma^\omega \eqdef \set{a_0a_1\dots \ST a_i \in \Sigma, i \in \N}$ is the set of streams composed of symbols from the finite set $\Sigma$. The set $\Sigma^\omega$ has continuum cardinality, and hence, can represent the reals and a class of distributions (Section~\ref{subsec:crev:real}). Once we \emph{represent} continuum-sized objects on a machine, we have an avenue for studying which functions are Type-2 computable. Throughout the rest of the chapter, we will abbreviate Type-2 computable as computable\footnote{Computability in the ordinary sense refers to Type-1 computability.} and use Type-2 computable for emphasis. We now review computable reals and distributions.

\subsection{Computability, Reals and Distributions}
\label{subsec:crev:real}

\paragraph{Computability and reals}
Intuitively, we can represent a real on a machine by encoding its binary expansion. More formally, we represent a real $x \in \R$ on a machine by encoding a fast Cauchy sequence of rationals that converges to $x$. Recall that a sequence $(q_n)_{n \in \N}$ where each $q_n \in \Q$ is \emph{Cauchy} if for every $\epsilon > 0$, there is an $N$ such that $|q_n - q_m| < \epsilon$ for every $n, m > N$. Thus, the elements of a Cauchy sequence become closer and closer to one another as we traverse the sequence. When $|q_n - q_{n + 1}| < 2^{-n}$ for all $n$, we call $(q_n)_{n \in \N}$ a \emph{fast Cauchy sequence}. Hence, the representation of a real as a fast Cauchy sequence evokes the idea of enumerating its binary expansion. A real $x \in \R$ is \emph{computable} if we can enumerate (uniformly in an enumeration of rationals) a fast Cauchy sequence that converges to $x$.

We give some examples of reals encoded as fast Cauchy sequences now.
\begin{example}[Rational] Consider two encodings of $0$ as a fast Cauchy sequence below.
\begin{itemize}[noitemsep,align=left]
\item[\emph{(Constant)}] Let $(x_n)_{n \in \N}$ where $x_n \eqdef 0$ for $n \in \N$.
\item[\emph{(Thrashing)}] Let $(y_n)_{n \in \N}$ where $y_n \eqdef \frac{1}{(-2)^{n + 1}}$ for $n \in \N$.
\end{itemize}
As $0$ itself is also a rational number, we can simply represent it as a constant $0$ sequence given by $(x_n)_{n \in \N}$. We can also represent $0$ as the sequence $(y_n)_{n \in \N}$, where the sequence jumps back and forth between positive and negative fractional powers of two as it converges towards $0$. $0$ is clearly a computable real.
\end{example}
\begin{example}[Irrational]
Let $x_n \eqdef 2 + \sum_{k=2}^{2+n} \frac{1}{k!}$. Then $(x_n)_{n \in \N}$ is a fast Cauchy encoding of $e$. It is easy to see that $e$ is a computable real.
\end{example}
\begin{example}[Non-computable]
Every real can be expressed as a fast Cauchy sequence so there are necessarily non-computable reals as well. Let $(M_n)_{n \in \N}$ be some enumeration of Turing machines. Let $t_0 \eqdef 1$ and
\[
t_{n+1} \eqdef \begin{cases}
2 \cdot t_n & \mbox{$M_n$ halts} \\
1 + t_n & \mbox{$M_n$ does not halt.}
\end{cases}
\]
Then $(x_n)_{n \in \N}$ where $x_n \eqdef \sum_{i=0}^n \frac{1}{2^{t_i}}$ is a fast Cauchy sequence that is not computable because the Halting problem is not decidable.
\end{example}

A function $f: \R \rightarrow \R$ is \emph{computable} if given a (fast Cauchy) sequence converging to $x \in \R$, there is an algorithm that outputs a (fast Cauchy) sequence converging to $f(x)$.\footnote{A function $f: \R^n \rightarrow \R$ is computable if given (fast Cauchy) sequences converging to $x_1, \dots, x_n \in \R$, there is an algorithm that outputs a (fast Cauchy) sequence converging to $f(x_1, \dots, x_n)$.} We emphasize that the algorithm must work generically for any input (fast Cauchy) sequence including the non-computable ones. we give some examples now.
\begin{example}[Addition]
The function $+_{0}: \R \rightarrow \R$ that adds $0$ is computable because an algorithm can obtain a (fast Cauchy) output sequence by adding the (fast Cauchy) input sequence element-wise to a (fast Cauchy) sequence of $0$.
\end{example}
\noindent Most familiar functions are computable (\eg, subtraction, multiplication, inverses, exponentiation, logarithms on non-negative reals, and trigonometric functions such as sines and cosines) so that there is an algorithm that transforms (fast Cauchy) inputs into (fast Cauchy) outputs. Nevertheless, there are familiar functions that are not computable.
\begin{example}[Non-computable]
Consider the function $=_0: \R \rightarrow \two$ that tests if the input is equal to $0$ or not. Intuitively, this function is not computable because we need to check the entire input sequence. For example, to check that the constant sequence is equivalent to the thrashing sequence, we have to check the entirety of both sequences, which cannot be done in finite time.
\end{example}

\paragraph{Computable metric spaces}
Topological spaces enable us to build a more general notion of computability on a space.\footnote{For more background on topology, we refer the reader to~\citet{munkres2000topology}.} For the purposes of introducing reals and distributions, we consider topological spaces with a notion of distance, \ie, \emph{metric spaces}. As a reminder, a \emph{metric space} $(X, d)$ is a set $X$ equipped with a \emph{metric} $d: X \times X \rightarrow \R$. A metric induces a collection of sets called \emph{(open) balls}, where a ball centered at $c \in X$ with radius $r \in R$ is the set of points within $r$ of $c$, \ie,~$B(c, r) \eqdef \set{x \in X \ST d(c, x) < r}$. The topology $\cO(X)$ associated with a metric space $X$ is the one induced by the collection of balls. Hence, the open balls of a metric space provide a notion of distance in addition to providing a notion of approximation.
\begin{example}
  $(\N, \discmet)$ endows the naturals $\N$ with the discrete topology (\ie, $\cO(\N) = 2^\N$), where $\discmet$ is the discrete metric (\ie, $d(n, m) \eqdef 0$ if $n = m$ and $d(n, m) \eqdef 1$ otherwise for $n, m \in \N$).
\end{example}
\begin{example}
  $(\R, \euclidmet)$ endows the reals $\R$ with the familiar Euclidean topology, where $\euclidmet$ is the standard Euclidean metric (\ie, $\euclidmet(x, y) \eqdef |x - y|$).
\end{example}
\begin{example}
  $(\two^\omega, \cantormet)$ endows the set of bit-streams $\two^\omega$ with the Cantor topology, where $\cantormet$ is defined as
  \[
  \cantormet(x, y) \eqdef \inf\set{\frac{1}{2^n} \ST x_n \neq y_n}. 
  \]
  One can check that a basic open set of the Cantor topology is of the form $a_1 \dots a_n\two^\omega \eqdef \set{b_1 b_2 \dots \in \two^\omega \ST b_i = a_i, 1 \leq i \leq n}$. That is, basic open sets of Cantor space fix finite-prefixes.
\end{example}

A computable metric space imposes additional conditions on a metric space so that a machine can enumerate successively more accurate approximations (according to the metric) of a point in the metric space. We need two additional definitions before we can state the definition. First, we say $S$ is \emph{dense} in $X$ if for every $x \in X$, there is a sequence $(s_n)_{n \in \N}$ that converges to $x$, where $s_n \in S$ for every $n$. Second, we say that $(X, d)$ is \emph{complete} if every Cauchy sequence comprised of elements from $X$ also converges to a point in $X$.
\begin{definition}
A \emph{computable metric space} is a tuple $(X, d, S)$ such that (1) $(X, d)$ is a complete metric space, (2) $S$ is a countable, enumerable, and dense subset, and (3) the real $d(s_i, s_j)$ is computable for $s_i, s_j \in S$~\citep[see][Def. 2.4.1]{hoyrup2009computability}. 
\label{def:crev:real:cms}
\end{definition}
\begin{example}
$(\R, \euclidmet, \Q)$ is a computable metric space for the reals where we use the rationals $\Q$ as the approximating elements. Note that we can equivalently use dyadic rationals as the approximating elements instead of $\Q$.
\label{ex:crev:real:real}
\end{example}

\paragraph{Computability and distributions}
A distribution over the computable metric space $(X, d, S)$ can be formulated as a point of the computable metric space
\[
(\cM(X), \prokhmet, \cD(S)) \,,
\]
where $\cM(X)$ is the set of Borel probability measures on a computable metric space $(X, d, S)$, $\prokhmet$ is the Prokhorov metric~\citep[see][Defn. 4.1.1]{hoyrup2009computability}, and $\cD(S)$ is the class of distributions with finite support at ideal points $S$ and rational masses~\citep[see][Prop. 4.1.1]{hoyrup2009computability}. The Prokhorov metric is defined as
\[
\prokhmet(\mu, \nu) \eqdef \inf \set{ \epsilon > 0 \ST \mu(A) \leq
  \nu(A^\epsilon) + \epsilon \mbox{ for every Borel $A$}} \;,
\]
where $A^\epsilon \eqdef \set{x \in X \ST d(x, y) < \epsilon \mbox{ for some } y \in A}$. One can check that the sequence below converges (with respect to the Prokhorov metric) to the (standard) uniform distribution $\cU(0, 1)$.
\[
\left\{0 \mapsto \frac{1}{2}, \frac{1}{2} \mapsto \frac{1}{2} \right\}, \left\{0
  \mapsto \frac{1}{4}, \frac{1}{4} \mapsto \frac{1}{4}, \frac{2}{4}
  \mapsto \frac{1}{4}, \frac{3}{4} \mapsto \frac{1}{4} \right\}, \dots,
\]
Thus, a uniform distribution can be seen as the limit of a sequence of increasingly finer discrete, uniform distributions. As with a computable real, we say that a distribution $\mu \in \cM(X)$ is \emph{computable} if we can enumerate (uniformly in an enumeration of a basis and rationals) a fast Cauchy sequence that converges to $\mu$.

Although the idea of constructing a (computable) distribution as a (computable) point is fairly intuitive for the standard uniform distribution, it may be more difficult to perform the construction for more complicated distributions. Fortunately, we can also think of a distribution on a computable metric space $(X, d, S)$ in terms of sampling, \ie, as a Type-2 computable function $\two^\omega \rightharpoonup X$. To make this more concrete, we sketch an algorithm that samples from the standard uniform distribution given a stream of fair coin flips. The idea is to generate a value that can be queried for more precision instead of a sample $x$ in its entirety.

Let $\distFair(a_1 \dots a_n \two^\omega) \eqdef 1/2^n$ be the distribution associated with a stream of fair coin flips where $0$ corresponds to heads and $1$ corresponds to tails. A sampling algorithm will interleave flipping coins with outputting an element to the desired precision, such that the sequence of outputs $(s_n)_{n \in \N}$ converges to a sample. For instance, one binary digit of precision for a standard uniform distribution corresponds to obtaining the point $1/2$ because it is within $1/2$ of any point in the unit interval. Demanding another digit of precision produces either $1/4$ or $3/4$ according to the result of a fair coin flip. This is encoded below using the function \texttt{bisect}\footnote{See the implementation of $\texttt{stdUniform}$ in Section~\ref{subsec:lib:ex} for the full definition.}, which recursively bisects an interval $n$ times, starting with $(0, 1)$, using the random bit-stream $\tape$ to select which interval to recurse on.
\begin{align*}
  \text{uniform} & : (\randty) \arrty (\natty \arrty \ratty) \\
  \text{uniform} & \eqdef \absexp{\tape}{\absexp{n}{ \texttt{bisect} \; \tape \; 0 \; 1 \; n }}
\end{align*}
In the limit, we obtain a single point corresponding to the sample.

The sampling view is (computably) equivalent to the view of a computable distribution as a point in an appropriate computable metric space. To state the equivalence, we need a few definitions. A \emph{computable
  probability space} $(X, \mu)$ is a pair where $X$ is a computable metric space and $\mu$ is a computable distribution~\citep[see][Def. 5.0.1]{hoyrup2009computability}. We call a distribution $\mu$ on $X$ \emph{samplable} if there is a computable function $s: (\two^\omega, \distFair) \rightharpoonup (X, \mu)$ such that $s$ is computable on $\dom(s)$ of full-measure (\ie, $\mu(X) = 1$) and is measure-preserving (\ie, $\mu = \distFair \circ s^{-1}$).
\begin{proposition}
  \emph{\citep[Computable iff samplable, see][Lem. 2 and Lem. 3]{freer2010posterior}\textbf{.}} A distribution $\mu \in \cM(X)$ on computable metric space $(X, d, \cS)$ is computable iff it is samplable.
\label{prop:bgprob:compiffsamp}
\end{proposition}
\noindent Thus we can equivalently specify computable distributions by writing sampling algorithms.

\section{A Library for Computable Distributions}
\label{sec:lib}

\emph{How do we implement continuous distributions as a library in a general-purpose programming language?} Our goal in this section is to translate the concepts about reals and distributions we saw previously in Section~\ref{sec:crev} into code. Towards this end, we sketch a Haskell library (Figure~\ref{fig:sem:lib}) that encodes reals and the sampling view of distributions.\footnote{The code is available at \url{https://github.com/danehuang/cdist-sketch}.} We emphasize that the library does not assume any reals, continuous distributions, or operations on them as black-box primitives.

\subsection{Library}
\label{subsec:lib:lib}

\begin{figure}[t]
\begin{hlisting}
module ApproxLib (Approx(..), CMetrizable(..), mkApprox, nthApprox) where

newtype Approx a = Approx { getApprox :: Nat -> a }

mkApprox :: (Nat -> a) -> Approx a  -- fast Cauchy sequence
nthApprox :: Approx a -> Nat -> a   -- project n-th approx.

class CMetrizable a where
    enum :: [a]                     -- countable, dense subset
    metric :: a -> a -> Approx Rat  -- computable metric
\end{hlisting}
\begin{hlisting}
module CompDistLib (RandBits, Samp(..), mkSamp) where
import ApproxLib
    
type RandBits = Nat -> Bool
newtype Samp a = Samp { getSamp :: RandBits -> a }

mkSamp :: (CMetrizable a) => (RandBits -> Approx a) -> Samp (Approx a)
mkSamp = Samp

instance Monad Samp where
  ...                               -- see text
\end{hlisting}
\caption{A Haskell library interface for expressing approximations in a computable metric space (module $\texttt{ApproxLib}$) and encoding (continuous) distributions (module $\texttt{CompDistLib}$). The library interface for reals (module $\texttt{RealLib}$) is not shown.}
\label{fig:sem:lib}
\end{figure}

The library consists of three modules. The first module $\texttt{ApproxLib}$ provides the interface for computable metric spaces. The second module $\texttt{RealLib}$ implements reals using the operations in $\texttt{ApproxLib}$ and the third module $\texttt{CompDistLib}$ implements (continuous) distributions. We go over the modules in turn now.

As we mentioned previously, the module $\texttt{ApproxLib}$ provides abstractions for expressing elements as a sequence of approximations in a computable metric space. The core type exposed by the module is $\approxty{\typ}$, which models an element of a computable metric space and can be read as an approximation by a sequence of values of type $\typ$. For example, a real can be given the type $\realty \eqdef \approxty{\ratty}$, meaning it is a sequence of rationals ($\ratty$) that converges to a real. We form values of type $\approxty{\typ}$ using $\approxlib :: (\natty \rightarrow \alpha) \rightarrow \approxty{\alpha}$, which requires us to check\footnote{We do not use the Haskell type system to enforce that the function to coerce contains a fast Cauchy sequence so the caller of $\approxlib$ needs to perform this check manually.} that the function we are coercing describes a fast Cauchy sequence, and project out approximations using $\anthlib :: \approxty{\alpha} \rightarrow \natty \rightarrow \alpha$.

In order to form the type $\approxty{\typ}$, values of type $\typ$ should support the operations required of a computable metric space. We can indicate the required operations using Haskell's type-class mechanism.
\begin{hlisting}
class CMetrizable a where
    enum :: [a]
    metric :: a -> a -> Approx Rat
\end{hlisting}
As a reminder, Haskell has lazy semantics so that the type $\texttt{[}\alpha\texttt{]}$ denotes a stream as opposed to a list. Thus $\texttt{enum}$ corresponds to an enumeration of type $\alpha$ where $\alpha$ is the type of the dense subset. When we implement an instance of $\metrizablecls{\typ}$, we should check that the implementation of $\texttt{enum}$ enumerates a dense subset and $\texttt{metric}$ computes a metric as a computable metric space requires (see Section~\ref{subsec:crev:real}).

Below, we give an instance of $\approxty{\ratty}$ for computable reals.
\begin{hlisting}
instance CMetrizable Rat where
    enum = 0 : [ toRational m / 2^n
               | n <- [1..]
               , m <- [-2^n * n..2^n * n]
               , odd m || abs m > 2^n * (n-1) ]
    metric x y = A (\_ -> abs (x - y))
\end{hlisting}
This instance enumerates the dyadic rationals, which are a dense subset of the reals. Note that there are many other choices here for the dense enumeration.\footnote{Algorithms that operate on computable metric spaces compute by enumeration so the algorithm is sensitive to the choice of enumeration.} In this instance, we can actually compute the metric as a dyadic rational, whereas a computable metric requires the weaker condition that we can compute the metric as a computable real.

Next, we can use the module $\texttt{ApproxLib}$ to implement computable operations on commonly used types. For example, a library for computable reals will contain the $\metrizablecls{\typ}$ instance implementation above and other computable functions. However, some operations are not realizable (\eg, equality of reals) and so this module does not contain all operations one may want to perform on reals (\eg, equality is defined on floats).
\begin{hlisting}
module RealLib (Real, pi, (+), ...) where
import ApproxLib

type Real = Approx Rat
instance CMetrizable Rat where
    ...

pi :: Real     
(+) :: Real -> Real -> Real
-- etc.
\end{hlisting}

The module $\texttt{CompDistLib}$ contains the implementation of distributions. A sampler $\sampty{\alpha}$ is a function from a bit-stream to values of type $\alpha$.\footnote{The type $\texttt{RandBits}$ is represented isomorphically as $\natty \rightarrow \boolty$ instead of $\texttt{[}\boolty\texttt{]}$.}
\begin{hlisting}
type RandBits = Nat -> Bool
newtype Samp a = Samp { getSamp :: RandBits -> a }
\end{hlisting}
We can implement an instance of the sampling monad as below.
\begin{hlisting}
instance Monad Samp where
    return x = Samp (const x)
    (>>=) s f = Samp ((uncurry (getSamp . f)) . (pair (getSamp s . fst) snd) . split)
        where pair f g = \x -> (f x, g x)
              split = pair even odd
              even u = (\n -> u (2 * n))
              odd u = (\n -> u (2 * n + 1))
\end{hlisting}
As expected, $\retkw$ corresponds to a constant sampler ($\texttt{const}$) that ignores its input randomness. The bind operator $\bindkw$ corresponds to a composition of samplers; we first split ($\texttt{split}$) the input randomness into two independent streams (via $\texttt{even}$ and $\texttt{odd}$), use one to sample from $\texttt{s}$, and continue with the other in $\texttt{f}$.

The module $\texttt{CompDistLib}$ provides the function $\samplerlib$ to coerce an arbitrary Haskell function of the appropriate type into a value of type $\sampty{\alpha}$.
\begin{hlisting}
mkSamp :: (CMetrizable a) => (RandBits -> Approx a) -> Samp (Approx a)
mkSamp = Samp
\end{hlisting}
We should call $\samplerlib$ only on sampling functions realizing Type-2 computable sampling algorithms.

\subsection{Examples}
\label{subsec:lib:ex}

We now encode discrete and continuous distributions using the constructs provided by library. These examples demonstrate how familiar distributions used in probabilistic modeling can be encoded in a Type-2 computable manner. As we walk through the examples, we will encounter some semantic issues that we would like a denotational semantics of probabilistic programs to handle. We will flag these in italics and revisit them after introducing a semantics for probabilistic programs (Section~\ref{sec:core}).

\paragraph{Discrete distribution}
Discrete distributions are much simpler compared to continuous distributions. Nevertheless, when paired with recursion, semantic issues do arise. For instance, consider the encoding of a geometric distribution with bias $1/2$, which returns the number of fair Bernoulli trials until a success. The distribution $\texttt{stdBernoulli}$ denotes a Bernoulli distribution with bias $1/2$.
\begin{hlisting}
stdGeometric :: Samp Nat
stdGeometric = do
  b <- stdBernoulli
  if b then return 1
       else stdGeometric >>= return . (\n -> n + 1)
\end{hlisting}
One possibility, although it occurs with zero probability, is for the draw from $\texttt{stdBernoulli}$ to always be false. Consequently, $\texttt{stdGeometric}$ diverges with probability zero. \emph{A semantics should clarify the criterion for divergence and show that this recursive encoding actually denotes a geometric distribution.}

\paragraph{Continuous distributions}
Next, we fill in the sketch of the standard uniform distribution we presented earlier. As a reminder, we need to convert a random bit-stream into a sequence of (dyadic) rational approximations.
\begin{hlisting}
stdUniform :: Samp Real
stdUniform = mkSamp (\u -> mkApprox (\n -> bisect (n+1) u 0 1 0))
    where
      bisect n u (l :: Rat) (r :: Rat) m
          | m < n && u m       =
              bisect n u l (midpt l r) (m+1)
          | m < n && not (u m) =
              bisect n u (midpt l r) r (m+1)
          | otherwise          =
              midpt l r
      midpt l r = l + (r - l) / 2
\end{hlisting}
The function \texttt{bisect} repeatedly bisects an interval specified by $(\texttt{l}, \texttt{r})$. By construction, the sampler produces a sequence of dyadic rationals. We can see that this sampling function is uniformly distributed because it inverts the binary expansion specified by the uniformly distributed input bit-stream. Once we have the standard uniform distribution, we can encode other primitive distributions (\eg,~normal, exponential, etc.) as transformations of the uniform distribution as in standard statistics using return and bind.

For example, we give an encoding of the standard normal distribution using the Marsaglia polar transformation.
\begin{hlisting}
stdNormal :: Samp Real
stdNormal = do
  u1 <- uniform (-1) 1
  u2 <- uniform (-1) 1
  let s = u1 * u1 + u2 * u2
  if s < 1 then return (u1 * sqrt (log s / s))
           else stdNormal
\end{hlisting}
The distribution $\uniDist{(-1)}{1}$ is the uniform distribution on the interval $(-1, 1)$ and can be encoded by shifting and scaling a draw from $\texttt{stdUniform}$. One subtle issue here concerns the semantics of $\texttt{<}$. As a reminder, equality on reals is not decidable. \emph{Consequently, although we have used $\texttt{<}$ at the type $\realty \rightarrow \realty \rightarrow \boolty$ in the example, it cannot have the standard semantics of deciding between $<$ and $\geq$.}

\paragraph{Singular distribution}
Next, we give an encoding of the Cantor distribution. The Cantor distribution is singular so it is not a mixture of a discrete component and a component with a density. Perhaps surprisingly, this distribution is computable. The distribution can be defined recursively. It starts by trisecting the unit interval, and placing half the mass on the leftmost interval and the other half on the rightmost interval, leaving no mass for the middle, continuing in the same manner with each remaining interval that has positive probability. We can encode the Cantor distribution by directly transforming a random bit-stream into a sequence of approximations.
\begin{hlisting}
cantor :: Samp Real
cantor = mkSamp (\u -> mkApprox (\n -> go u 0 1 0 n))
    where 
      go u (left :: Rat) (right :: Rat) n m 
        | n < m && u n       =
            go u left (left + pow) (n + 1) m
        | n < m && not (u n) =
            go u (right - pow) right (n + 1) m
        | otherwise          =
            right - (1 / 2) * pow
          where pow = 3 ^^ (-n) 
\end{hlisting}
The sampling algorithm keeps track of which interval it is currently in specified by \texttt{left} and \texttt{right}. If the current bit is $1$, we trisect the left interval. Otherwise, we trisect the rightmost interval. The number of trisections is bounded by the precision we would like to generate the sample to. \emph{Crucially, the encoding makes use of the idea of generating a sample to arbitrary accuracy using a representation instead of the sample in its entirety.}

\paragraph{Partiality and distributions}
The next series of examples explores issues concerning distributions and partiality.
\begin{hlisting}
botSamp :: (CMetrizable a) => Samp (Approx a)
botSamp = botSamp

botSampBot :: (CMetrizable a) => Samp (Approx a)
botSampBot = mkSamp (\_ -> bot)
    where bot = bot
\end{hlisting}
In the term $\texttt{botSamp}$, we define an infinite loop at the type of samplers. Intuitively, this corresponds to the case where we fail to provide a sampler, \ie, an error in the worst possible way. In the term $\texttt{botSampBot}$, we produce a sampler that fails to generate a sample to any precision. In other words, we provide a sampler that is faulty in the worst possible way. We can try to observe the differences in the implementation (if any).

\begin{minipage}{0.45\textwidth}
\begin{hlisting}
alwaysDiv :: Samp Real
alwaysDiv = do
  _ <- botSamp ::
         Samp Real
  stdUniform
\end{hlisting}
\end{minipage}
\begin{minipage}{0.45\textwidth}
\begin{hlisting}
neverDiv :: Samp Real
neverDiv = do
  _ <- botSampBot ::
         Samp Real
  stdUniform
\end{hlisting}
\end{minipage}

\noindent If we run the term $\texttt{alwaysDiv}$ on the left, we will see that the program always diverges. When we run the term $\texttt{neverDiv}$ on the right, we will draw from the sampler $\texttt{botSampBot}$ but discard the result. Due to Haskell's lazy semantics, this computation is ignored and the entire term behaves as a standard uniform distribution. \emph{We would like a denotational semantics to reflect the differences in the operational behavior between these two terms.}

\paragraph{Commutativity and independence}
We end by considering the difference between a sampling and distributional interpretation of probabilistic programs. Below, we give equivalent encodings of distributions by commuting the order of sampling from independent distributions, but leaving everything else fixed.

\begin{minipage}{0.45\textwidth}
\begin{hlisting}
myNormal :: Samp Real
myNormal = do
  x <- normal (-1) 1 
  y <- normal (1) 1
  return (x + y)
\end{hlisting}
\end{minipage}
\begin{minipage}{0.45\textwidth}
\begin{hlisting}
myNormal' :: Samp Real
myNormal' = do
  y <- normal (1) 1
  x <- normal (-1) 1 
  return (x + y)
\end{hlisting}
\end{minipage}

\noindent From a sampling perspective, the two distributions are not strictly equivalent because the stream of random bits is consumed in a different order; consequently, the samples produced by \texttt{myNormal} and \texttt{myNormal'} may be different. \emph{Thus, while a sampling semantics is easily implementable, we would also like a distributional semantics to enable reasoning about the distributional equivalence of programs. For instance, this would enable us to reason that two different sampling algorithms for the same distribution are equivalent.}

\subsection{Notes}
\label{subsec:lib:notes}

The implementation we have sketched is a proof of concept that shows that we can realize the interface by implementing computable distributions and operations on them as Haskell code. We note that there are multiple approaches to coding up Type-2 computability as a library. One prominent alternative is given by synthetic topology~\citep{escardo2004synthetic}, which assumes that the function space in the programming language used to code up topological results is continuous and derives the notion of an open set. These ideas can be used to help us structure an implementation.

One shortcoming of the library, and implementations of Type-2 computability more generally, is efficiency. We intend the presentation of the library as a means to sketch the connection of the computation with the mathematics. In practice, there are still reasons for using floating point arithmetic. First, inference algorithms are computationally intensive, even assuming operations on reals and distributions are constant-time, so one is willing to make tradeoffs for efficiency. Second, it is not necessary to compute answers to arbitrary accuracy for most applications. Notably, most inference algorithms already make approximations as the solutions to many interesting models are analytically intractable. Thus, there is still a (large) gap in practice between semantics and implementation. For ideas on how to implement Type-2 computability efficiently, we refer the reader to~\citet{bauer2008implementing} and~\citet{lambov2007reallib}.

Lastly, in our description of the library, we have elided one important detail. One computable function we need to encode is the {\em modulus} of a computable function between computable metric spaces. The modulus $g: (X \rightarrow Y) \rightarrow \N \rightarrow \N$ of a computable function $f: X \rightarrow Y$ between computable metric spaces $(X, d_X, \cS_X)$ and $(Y, d_Y, \cS_Y)$ is a function that computes the number of input approximations consumed to produce an output approximation to a specified precision. For example, if the algorithm realizing $f$ looks at $s_{i_0}^X, \hdots, s_{i_{41}}^X$ to compute an output $s_{i_n}^Y$ such that $d_Y(s_{i_n}^Y, f(x)) < 2^{-(n + 1)}$ and $(s_{i_m}^X)_{m \in \N} \to x$, then the modulus $g(f)(n)$ is $42$. Within a machine model, one can simply ``look at the tape and head location" to obtain the modulus. However, one can show that the modulus of continuity is not expressible in a functionally-extensional language. This in essence follows from the fact that the modulus of two extensionally equivalent functions may not be equivalent. We can use Haskell's imprecise exceptions mechanism~\citep[see][]{peyton1999semantics}, an impure feature, in a restricted manner to express the modulus.\footnote{See \url{http://math.andrej.com/2006/03/27/sometimes-all-functions-are-continuous}.}
\newcommand\cp{$+$} 
\newcommand\cm{$\checkmark$}

\section{Mathematical Structures for Modeling the Library}
\label{sec:math}

\emph{What mathematical structures can we use to model such a library?} Now that we have seen that we can implement reals and continuous distributions in code, our next task is to find mathematical structures that can be used to faithfully model the implementation. In doing so, we will set ourselves up for giving \emph{denotational semantics} to probabilistic programs under the additional constraint that the model takes computability into account (Section~\ref{sec:core}).

Towards this end, we review \emph{topological domains}, an alternative to traditional domain theory (Section~\ref{subsec:math:dom}). Topological domains support all the standard domain-theoretic constructions needed to model PCF-like languages as well as capture the notion of Type-2 computability, and hence, can form the basis of a semantics for PCF-like languages. Next, we encode distributions as topological domains. We do this for a sampling view (Section~\ref{subsec:math:sampling}) and a distributional view (Section~\ref{subsec:math:valuations}) based on \emph{valuations}, a topological variant of a measure. We also construct a probability monad~\citep{giry1982categorical} on countably based topological (pre)domains, which includes computable metric spaces, so we can model the monadic implementation of distributions in the library. 

Finally, we put the approach proposed here, which emphasizes Type-2 computability, in perspective. We begin by exploring an alternative approach to capturing Type-2 computability via \emph{realizability} (Section~\ref{subsec:math:realize}). Roughly speaking, we can view a constructive logic as a ``programming language" that we can use to program computable distributions. We end by reviewing alternative structures that can be used to model the semantics of probabilistic programs (Section~\ref{subsec:math:alt}).

\subsection{Domains and Type-2 Computability}
\label{subsec:math:dom}

In this section, we review \emph{topological domains}. Unlike a CPO, a topological domain in general does not carry the Scott topology, and hence, does not consider the partial order primary. Instead, topological domains start with the topology as primary and derive the order. For a complete treatment, we refer the reader to~\citet{battenfeld2008td} and the references within~\citep[\eg, see][]{battenfeld2004category,battenfeld2006compactly,battenfeld2007convenient}. Towards this end, we will follow the overview given by~\citet{battenfeld2007convenient} to introduce the main ideas, which constructs topological domains in two steps: (1) connecting computability to topology and (2) relating topology to order. Most of this overview can be skimmed upon a first read, although the examples will be helpful. At the end, we will summarize the relevant structure that makes topological domains good candidates for modeling probabilistic programs. In Section~\ref{sec:core}, we will use this structure to give semantics to a core language.

\paragraph{Computability to topology}
Topological domain theory starts with the observation that topological spaces provide a good model of \emph{datatypes}. In short, a point in a topological space corresponds to an inhabitant of a datatype and the open sets of the topology describe the observable properties of points. Consequently, one can test if an inhabitant of a datatype satisfies an observable property by performing a (potentially diverging) computation that tests if the point is contained in an open set. To make use of this observation, topological domain theory builds off of the Cartesian closed category of $qcb_0$ spaces\footnote{$qcb_0$ stands for a $T_0$ quotient of a countably based space.}~\citep[\eg, see][]{escardo2004comparing}, a subcategory of topological spaces that makes the connection between computation and topology precise. It is helpful to introduce a $qcb_0$ space by way of a \emph{represented space} which starts with the idea of realizing computations on a machine model before adding back the topological structure.

\begin{definition}
A \emph{represented space} $(X, \rep_X)$ is a pair of a set $X$ with a partial surjective function $\rep_X : \two^\omega \rightharpoonup X$ called a \emph{representation}.
\end{definition}
\noindent We call $p \in \two^\omega$ a \emph{name} of $x$ when $\rep_X(p) = x$. Thus, a name encodes an element of the base set $X$ as a bit-stream which in turn can be computed on by a Turing machine. A \emph{realizer} for a function $f : (X, \rep_X) \rightarrow (Y,\rep_Y)$ is a (partial) function $F : \two^\omega \rightharpoonup \two^\omega$ such that $\rep_Y(F(p)) = f(\rep_X(p))$ for $p \in \dom(f \circ \rep_X)$.  A function $f : X \rightarrow Y$ between represented spaces is called \emph{computable} if it has a computable realizer. It is called \emph{continuous} if it has a continuous realizer (with respect to the Cantor topology).\footnote{Note that a continuous function $f : X \rightarrow Y$ between represented spaces does not mean that $f: X \rightarrow Y$ is a topologically continuous function with respect to the final topologies induced by the respective representations.} Unfolding the definition of continuity of a (partial) function $f : \two^\omega \rightharpoonup \two^\omega$ on Cantor space shows that it encodes a \emph{finite prefix property}---this means that a machine can compute $f(p)$ to arbitrary precision after consuming a finite amount of bits of $p$ in finite time when $f$ is continuous.

In order to relate the machine-model view to a topology so we can define a $qcb_0$ space, we will need a notion of an \emph{admissible representation}. A representation $\rep_X$ of $X$ is \emph{admissible} if for any other representation $\rep_X'$ of $X$, the identify function on $X$ has a continuous realizer~\citep[][Defn. 3.10]{battenfeld2007convenient}.
\begin{definition}
A \emph{$qcb_0$ space} is a represented space $(X, \rep_X)$ with \emph{admissible representation} $\rep_X$. 
\end{definition}
\noindent The topology is the \emph{quotient topology} (or \emph{final topology}) induced by the representation $\rep_X$. If $X$ and $Y$ are $qcb_0$ spaces, then the topologically continuous functions between them coincide with those that have continuous realizers~\citep[][Cor. 3.13]{battenfeld2007convenient}, which gives the same characterization as an admissible represented space. We give two examples of $qcb_0$ spaces to illustrate the corresponding realizers and topologies.
\begin{example}
Define the set $\bbS \eqdef \set{\bot, \top}$ with representation $\rep_\bbS(\bot) \eqdef 00\dots$ and $\rep_\bbS(\top) \eqdef p$ for $p \neq 00\dots$. Then $(\bbS, \rep_\bbS)$ is a $qcb_0$ space known as \emph{Sierpinski} space. In particular, Sierpinski space encodes the notion of semi-decidability---a Turing machine semi-decides that a proposition holds (encoded as $\top$) only if it eventually outputs a non-zero bit.
\end{example}
\begin{example}
Let $(X, d, S)$ be a computable metric space. Then $(X, \rep_{\text{Metric}})$ is a $qcb_0$ space with admissible representation $\rep_{\text{metric}}$ that uses fast Cauchy sequences as names. More concretely, $(\rep_\Q(w_n))_{n \in \N} \to \rep_{\text{metric}}(p)$ where $\rep(p) = \cpair{w_1, w_2, \dots}$. As a special case, $(\R, \rep_\R)$ is a represented space, where $\rep_\R$ is a representation that uses fast Cauchy sequences of rationals as names.
\end{example}

\paragraph{Topology to order}
The next piece of structure topological domain theory imposes is the order-theoretic aspect. The idea is to use the standard interpretation of recursive functions as the least upper bound of an ascending chain of the approximate functions obtained by unfolding. Because topological domain theory takes the topology as primary and the order as secondary, this task requires some additional work.

Recall that we can convert a topological space into a preordered set via the \emph{specialization preorder}, which orders $x \ord y$ if every open set that contains $x$ also contains $y$. We write $\spec$ to convert a topological space into a preordered set. Intuitively, $x \ord y$ if $x$ contains less information than $y$. For a metric space, we can always find an open ball that separates two distinct points $x$ and $y$ (because the distance between two distinct points is positive). Hence, the specialization preorder of a metric space always gives the discrete order (\ie, information ordering), and hence degenerately, a CPO.

\begin{definition}
\citep[][Defn. 5.1]{battenfeld2007convenient}\textbf{.} A $qcb_0$ space is called a \emph{topological predomain} if every ascending chain $(x_i)_{i \in \N}$ (with respect to the specialization preorder $\ord$) has an upper bound $x$ such that $(x_i)_{i \in \N} \rightarrow x$ (with respect to its topology). 
\end{definition}
\noindent Thus, we see in the definition that a topological predomain (1) builds off of a $qcb_0$ space and (2) ensures that least upper bounds of increasing chains exist. The former condition provides the topology and theory of effectivity while the latter condition prepares us for modeling least fixed-points. The following provides a useful characterization of $qcb_0$ spaces that relates the topology back to the order. 
\begin{definition}
  \citep[][Defn. 5.3]{battenfeld2007convenient}\textbf{.} A topological space $(X, \cO(X))$ is a \emph{monotone convergence space} if its specialization order is a CPO and every open is Scott open.
\end{definition}
\begin{proposition}
  \emph{\citep[][Prop. 5.4]{battenfeld2007convenient}\textbf{.}} A $qcb_0$ space is a topological predomain iff it is a monotone convergence space.
\end{proposition}
\noindent Hence, we see that the Scott topology is in general finer than the topology associated with a topological predomain.

Analogous to standard domain theory, a topological predomain is called a \emph{topological domain} if it has least element, written $\bot$, under its specialization order~\citep[][Defn. 5.6]{battenfeld2007convenient}.
\begin{proposition}
\emph{\citep[][Thm. 5.7]{battenfeld2007convenient}\textbf{.}}
Every continuous endofunction on a topological domain has a least fixed-point.
\end{proposition}

We look at the relation between order and topology more closely through a series of examples below.
\begin{example}
Consider the \emph{discrete} CPO $(\N, \ord_{\text{discrete}})$ with \emph{discrete ordering} $\ord_{\text{discrete}}$, (\ie, $n \ord_{\text{discrete}} m$ if $n = m$). The Scott topology on this CPO gives the \emph{discrete topology}, \ie, $\cO(\N) = \set{\set{n} \ST n \in \N}$. The specialization preorder applied to the resulting topology gives back the original CPO. Thus, we additionally see that the topological predomain coincides with the CPO.
\label{ex:math:dom:nat}
\end{example}
\begin{example}
Consider the CPO $(\set{[a, 1) \ST a \in \R} \cup \set{[0, 1]}, \subseteq)$ with ordering given by set inclusion. The Scott topology on this CPO gives the \emph{lower topology}, \ie, $\cO([0, 1]) = \set{(a, 1] \ST a \in [0, 1) } \cup \set{[0, 1]}$. Like the previous example, the specialization preorder applied to the resulting topology gives back the original CPO. Hence, the topological domain also coincides with the CPO.
\label{ex:math:dom:lower}
\end{example}
\noindent In the two examples above, we saw instances where the order and topology coincide. In the next two examples, we will see cases where they differ, thus highlighting differences between CPOs and topological (pre)domains.
\begin{example}
The reals $\R$ with Euclidean topology is a metric space, and hence, the specialization preorder gives a discrete CPO $(\R, \ord_{\text{discrete}})$. However, the Scott topology of the resulting discrete CPO is the discrete topology. Hence, the topologies do not coincide.
\end{example}
\begin{example}
The Scott continuous functions from $\R$ to $\R$ contain all functions. However, the space of functions between the topological predomains $\R$ and $\R$ contain just the continuous ones.
\end{example}
\noindent The last example concerns modeling divergence for reals.
\begin{example}
The \emph{partial reals} $\tilde{\R}$~\citep[\eg, see][]{escardo1996pcf} can be modeled as (closed) intervals $[l, u]$ ordered by reverse inclusion where $l$ is a lower-real and a $u$ is an upper-real. The subspace of the maximal elements yields the familiar Euclidean topology. Note that $\tilde{\R}_\bot \neq \R_\bot$.
\label{ex:math:dom:partialreal}
\end{example}

\begin{figure}[t]
\centering
\begin{tabular}{c|c|c|c|c|c|c|c}
Construction & $D \proddom[] E$ & $D \arrdom[] E$ & $D \sumdom[] E$ & $D \sproddom[] E$ & $D \sarrdom[] E$ & $D \ssumdom[] E$ & $\liftdom{D}$ \\ \hline
\textbf{TP} & \cm\cp & \cm\cp & \cm\cp & & & & \cm \\
\textbf{TD} & \cm\cp & \cm\cp & & \cm & \cm & \cm & \cm \\
\textbf{TD}$_{!}$ & \cm\cp & \cm & & \cm & \cm\cp & \cm\cp & \cm
\end{tabular}
\caption{Summary of constructs on topological predomains (category \textbf{TP}), topological domains (category \textbf{TD}), and topological domains with strict morphisms (category \textbf{TD}$_{!}$). (Compare this figure with one for CPOs~\citep[][pg. 46]{abramsky1994domain}.) The symbol $\checkmark$ indicates that the category is closed under that construct and the symbol $+$ additionally indicates that it corresponds to the appropriate categorical construct.}
\label{fig:math:dom:td}
\end{figure}

\paragraph{Categorical structure}
We end by summarizing the categorical structure of topological domains (Figure~\ref{fig:math:dom:td}) applicable to giving semantics to probabilistic programs.\footnote{We include sums ($D \sumdom[] E$) and coalesced sums ($D \ssumdom[] E$) for completeness. Similar to a smash product, a coalesced sum $D \ssumdom[] E$ identifies the least element of $D$ with the least element of $E$.} In short, topological (pre)domains possess essentially the same categorical structure as their CPO counterparts. Hence, we will be able to give semantics to programming languages using topological domains in much the same way that we use CPOs. 

The relevant categories include $\textbf{TP}$ (topological predomains and continuous functions),\footnote{\textbf{TP} is a full reflective exponential ideal of \textbf{QCB} (category with $qcb_0$ spaces as objects and continuous functions as morphisms)~\citep[][Thm. 5.5]{battenfeld2007convenient}.} $\textbf{TD}$ (topological domains and continuous functions),\footnote{\textbf{TD} is an exponential ideal of \textbf{QCB} and is closed under countable products in \textbf{QCB}~\citep[][Thm. 5.9]{battenfeld2007convenient}.} and $\textbf{TD}_{!}$ (topological domains and strict continuous functions).\footnote{\textbf{TD}$_!$ (1) is countably complete (limits inherited from \textbf{QCB}), (2) has countable coproducts, and (3) $\ssumdom[]$ and $\sarrdom[]$ (with $\mathbb{S}$ as unit) provides symmetry monoidal closed structure on \textbf{TD}$_!$~\citep[][Thm. 6.1, Thm. 6.2, Prop. 6.4]{battenfeld2007convenient}.} We will use the notation below for categorical constructions with the usual semantics.
\begin{itemize}[noitemsep,align=left]
    \item[(\emph{Function})] We write $D \arrdom[] E$ for continuous functions ($D \sarrdom[] E$ for strict continuous functions); the corresponding operation includes $\cateval: (D \arrdom[] E) \proddom[] D \arrdom[] E$, $\catuncurry: (D \arrdom[] E \arrdom[] F) \arrdom (D \proddom[] E \arrdom F)$, and $\catcurry: (D \proddom[] E \arrdom F) \arrdom[] D \arrdom[] E \arrdom[] F$. We will subscript function space $\arrdom[]$ with the appropriate category when it is not clear from context which function space we are referring to, \eg, $D \arrdom[\textbf{TD}] E$.
    \item[(\emph{Product})] We write $D \proddom[] E$ for products ($D \sproddom[] E$ for smash products);\footnote{A smash product $D \sproddom[] E$ identifies the least element of $D$ with the least element of $E$.} the corresponding operations include first projection $\pi_1: D \proddom[] E \arrdom[] D$, second projection $\pi_2: D \proddom[] E \arrdom[] E$, and pairing $\langle \cdot, \cdot \rangle: (D \arrdom[] E) \proddom[] (D \arrdom[] F) \arrdom[] (D \arrdom[] E \proddom[] F)$.
    \item[(\emph{Lift})] $\liftdom{D}$ lifts a \emph{(pre)domain}; the corresponding operations include lifting elements $\lift{\cdot}: D \arrdom \liftdom{D}$, lifting the domain of a function $\liftd: (D \arrdom[] \liftdom{E}) \arrdom[] (\liftdom{D} \arrdom[] \liftdom{E})$, lifting the codomain of a function $\liftc: (D \arrdom[] E) \arrdom[] (D \arrdom[] \liftdom{E})$, and unlifting elements $\unlift{\cdot}: \liftdom{D} \arrdom D$ for $D$ ($\unlift{\lift{d}} = d$ and undefined otherwise). Given a morphism $f: D \arrdom[] E$, we write $f_\bot: \liftdom{D} \arrdom[] \liftdom{E}$ to refer to the morphism with lifted domain and codomain.
\end{itemize}

\subsection{Sampling}
\label{subsec:math:sampling}

As a reminder, the library implementation converts an input bit-stream into a sample in the desired space. Hence, we begin by encoding the sampling implementation of distributions from the library as a topological domain.

Define an (endo)functor $\sampfunc$ that sends a topological predomain $D$ to a sampler on $D$ and a morphism to one that composes with the underlying sampler. Then, the topological domain $\sampfunc(D)$ is a sampler producing values in the lifted topological domain $\liftdom{D}$.
\begin{proposition}
The functor $\sampfunc$ defined as
\begin{align*}
  \sampfunc(D : \textbf{TP}) & \eqdef \two^\omega \arrdom[] \liftdom{D} \\
  \sampfunc(f : D \arrdom[] E) & \eqdef \absD{s}{f_\bot \circ s} \,,
\end{align*}
is well-defined, where $\two^\omega$ is the topological predomain equipped with the Cantor topology.
\end{proposition}
\noindent The least element is one that maps all bit-streams to $\bot$. Next, we define three operations on samplers. The first operation creates a sampler that ignores its input bit-randomness and always returns $d$:
\begin{align*}
\detfn : D & \arrdom[] \sampfunc(D) \\
\detfn(d) & \eqdef \catconst(\lift{d})
\end{align*}
where $\catconst: D \arrdom[] (E \arrdom D)$ produces a constant function. 

The second operation splits an input bit-stream $u$ into the bit-streams indexed by the even indices $u_e$ and the odd indices $u_o$:
\begin{align*}
\splitfn : \two^\omega & \arrdom[] \two^\omega \proddom[] \two^\omega \\
\splitfn(u) & \eqdef (u_e, u_o) \,.
\end{align*}
Note that if $u$ is a sequence of independent and identically distributed bits, then both $u_e$ and $u_o$ will be as well.

The third operation sequences two samplers:
\begin{align*}
\sampfn : \sampfunc(D) & \proddom[] (D \arrdom[] \sampfunc(E)) \arrdom[] \sampfunc(E) \\
\sampfn(s, f) & \eqdef \catuncurry(\liftd(f)) \circ \langle s \circ \pi_1, \pi_2 \rangle \circ \splitfn .
\end{align*}
It splits the input bit-randomness and runs the sampler $s$ on one of the bit-streams obtained by splitting to produce a value. That value is fed to $f$, which in turn produces a sampler that is run on the other bit-stream obtained by splitting.

\subsection{Valuations and a Probability Monad}
\label{subsec:math:valuations}

Our goal now is encode distributions as \emph{valuations} in the framework of topological domains. Once we have done so, we can interpret distribution terms in the library as elements of the appropriate topological domain. Next, we define the probability monad, which will be restricted to countably based topological (pre)domains. Consequently, the probability monad in $\langprob$ will be restricted to distributions on countably based spaces, which includes commonly used spaces such as reals and products of countably based spaces (Section~\ref{sec:core}).

\paragraph{Valuations and measures}
A valuation shares many of the same properties as a measure, and hence, can be seen as a topological variation of distribution.
\begin{definition}
A \emph{valuation} $\nu: \cO(X) \rightarrow [0, 1]$ is a function that assigns to each open set of a topological space $X$ a probability such that it is (1) strict ($\nu(\emptyset) = 0$), (2) monotone ($\nu(U) \leq \nu(V)$ for $U \subseteq V$), and (3) modular ($\nu(U) + \nu(V) = \nu(U \cup V) + \nu(U \cap V)$ for every open $U$ and $V$).
\end{definition}
\noindent One key difference between valuations and measures is that valuations are not required to satisfy countable additivity. Indeed, countable additivity is perhaps one of the defining features of a measure. We can rectify this situation for valuations by restricting attention to the \emph{$\omega$-continuous valuations}. As a reminder, a valuation $\nu$ is called \emph{$\omega$-continuous} if $\nu(\bigcup_{n \in \N} V_n) = \sup_{n \in \N} \nu (V_n)$ for $(V_n)_{n \in \N}$ an increasing sequence of opens. Hence, the countable additivity of $\mu$ encodes the $\omega$-continuous property. Importantly, note that every Borel measure $\mu$ can be restricted to the lattice of opens, written $\mu|_{\cO(X)}$, resulting in an $\omega$-continuous valuation. Every Borel measure $\mu$ on $X$ can be restricted to an $\omega$-continuous valuation $\mu|_{\cO(X)} : [\openpo{X} \arrdom[\CPO] \zopo]$~\citep[see][Sec. 3.1]{schroder2007admissible}. Moreover, $\mu$ is uniquely determined by its restriction to the opens $\mu|_{\cO(X)}$.\footnote{Note that the $\omega$-continuous condition encodes what it means for a function to be $\omega$-Scott continuous, \ie, an $\omega$-CPO continuous function.} In other words, we can identify distributions on topological spaces with $\omega$-continuous valuations.

\paragraph{Encoding valuations}
The presence of topological and order-theoretic structure suggests two strategies for encoding valuations as topological domains. In the first approach, we would take a realizer point of view as every topological domain is also a $qcb_0$ space. Under this approach, we would (1) define an admissible representation of the space of opens $\cO(X)$, (2) define an admissible representation of the interval $[0, 1]$, and (3) verify that a representation of a valuation $\cO(X) \rightarrow [0, 1]$ using the canonical function space representation is admissible and properly encodes a valuation. In the second approach, we would take an order-theoretic point of view. Under this approach, we would (1) verify that the space of opens $\cO(X)$ is a topological domain, (2) verify that the interval $[0, 1]$ is a topological domain, and (3) verify that the continuous functions $\cO(X) \arrdom[] [0, 1]$ encodes a valuation correctly. In either strategy, a common thread is that we need to encode the opens $\cO(X)$ and the interval $[0, 1]$. We start with the realizer perspective.

Let $\sC(X, \mathbb{S})$ be the space of continuous functions between the represented spaces $X$ and $\mathbb{S}$. Let $[0, 1]_< \eqdef ([0, 1], \rep_<)$ be the represented space with representation $\rep_<$ that represents $r \in [0, 1]$ as all the rational lower bounds. Next, we define the opens $\cO(X)$ and the interval $[0, 1]$ for the order-theoretic perspective. Let $\openpo{X} \eqdef (\cO(X), \subseteq)$ be the lattice of opens (and hence a CPO) of a topological space $X$ ordered by subset inclusion. Let $\zopo \eqdef ([0, 1], \leq)$ be the interval $[0, 1]$ ordered by $\leq$. The next proposition shows that the realizer perspective and the order-theoretic perspective are equivalent.
\begin{proposition}\hfill
\begin{enumerate}[noitemsep]
\item $[0, 1]_< \cong \zopo$ and
\item $\sC(X, \mathbb{S}) \cong \openpo{X}$ when $X$ is an admissible represented space.\footnote{The second item is due to~\citet[Thm. 3.3]{schroder2007admissible}.}
\end{enumerate}
\end{proposition}
\noindent 
The next proposition shows that the realizer and order-theoretic views are equivalent under the additional assumption that the base topological space is countably based.
\begin{proposition} Let $(X, \cO(X))$ be a countably based topological space.\hfill
\begin{enumerate}[noitemsep]
\item $[\openpo{X} \arrdom[\CPO] \zopo] \cong \sC(\cO(X), \zorep)$ and
\item $[\openpo{X} \arrdom[\CPO] \zopo] \cong [\openpo{X} \arrdom[\TD] [0, 1]^\uparrow]$.\footnote{The first item is due to~\citet[Sec. 3.1, Thm 3.5, Cor. 3.5]{schroder2007admissible}. For the second item, recall that every $\omega$-continuous pointed CPO with its Scott topology coincides with a topological domain~\citep{battenfeld2007convenient}. The least element is the valuation that maps every open set to $0$.}
\end{enumerate}
\label{prop:math:valuations:valprop}
\end{proposition}
\noindent Proposition~\ref{prop:math:valuations:valprop} gives three equivalent views of a valuation as (1) a CPO continuous function, (2) a continuous map between represented spaces, and (3) a continuous function between topological domains. View (2) indicates that there is an associated theory of effectivity on valuations. We will use this view to give semantics to probabilistic programs.

\paragraph{Integration}
Similar to how one can integrate a measurable function with respect to a measure, one can integrate a lower semi-continuous function with respect to a valuation. Let $X$ be a represented space and $\mu \in \cM_1(X)$ where $\cM_1(X)$ is the collection of Borel measures on $X$ that have total measure $1$.
\begin{proposition}
The integral of a lower semi-continuous function $f \in \sC(X, \zorep)$ with respect to a Borel measure $\mu$
\[
\int : \sC(X, \zorep) \times \cM_1(X) \rightarrow \zorep
\]
is lower semi-continuous~\citep[see][Prop. 3.6]{schroder2007admissible}. In fact, it is even lower semi-computable~\citep[][Prop. 3.6]{schroder2007admissible}~\citep[][Prop. 4.3.1]{hoyrup2009computability}.
\end{proposition}
\label{prop:sem:valuations:int}
\noindent The integral is defined in an analogous manner to the Lebesgue integral, \ie, as the limit of step functions on opens instead of measurable sets. The integral possesses many of the same properties, including Fubini and monotone convergence.

\paragraph{Probability monad}
Finally, we combine the results about valuations and integration to define a probability monad. Let $\textbf{TP}_\omega$ be the full subcategory of $\textbf{TP}$ where the objects are countably based. Define the (endo)functor $\probfunc$ on countably based topological predomains that sends an object $D$ to the space of valuations on $D$ and a morphism to one that computes the pushforward.
\begin{proposition}
  The functor $\probfunc$ defined as
\begin{align*}
  \probfunc(D: \textbf{TP}_\omega) & \eqdef \openpo{D} \arrdom[] \zopo \\
  \probfunc(f: D \arrdom[] E) & \eqdef \absD{\mu}{\mu \circ f^{-1}}
\end{align*}
is well-defined.
\end{proposition}
\noindent It is straightforward to check that $\probfunc$ is a functor. We can construct a probability monad using the functor $\probfunc$.
\begin{proposition}
The triple $(\probfunc, \mret, \mbind)$ is a monad, where
\begin{align*}
  \mret(x)(U) & \eqdef \indexp{U}{x} \\
  (\mu \mbind f)(U) & \eqdef \int f_U \; d\mu \mbox{  where $f_U(x) = f(x)(U)$.}
\end{align*}
\end{proposition}
\noindent It is largely straightforward to check that $(\probfunc, \mret, \mbind)$ is a monad.\footnote{In the case of bind, we can check that the identities involving integrals holds via standard arguments~\citep[\eg, see][]{jones1989thesis}.}

\subsection{Realizability}
\label{subsec:math:realize}

\newcommand\IntTy{\texttt{Int}}
\newcommand\NonNegRatTy{\texttt{NonNegRat}}
\newcommand\RatTy{\texttt{Rat}}
\newcommand\NatTy{\texttt{Nat}}
\newcommand\BasisTy{\texttt{Basis}}
\newcommand\BaireTy{\texttt{Baire}}

\newcommand\synlarr{\leftarrow}
\newcommand\synrarr{\rightarrow}
\newcommand\syniff{\leftrightarrow}
\newcommand\synand{\,\land\,}

\newcommand\sierp{\mathbb{S}}
\newcommand\baire{\mathbb{B}}
\newcommand*\quot[2]{{^{\textstyle #1}\big/_{\textstyle #2}}}
\newcommand\LR{\R_<}
\newcommand\UR{\R_>}
\newcommand\relmv{\sim}

Sections~\ref{subsec:math:dom},~\ref{subsec:math:sampling}, and~\ref{subsec:math:valuations} taken together provide enough structure for giving semantics to probabilistic programs with continuous distributions. Thus, the reader interested in seeing the semantics ``in action" in a core language can skip ahead to Section~\ref{sec:core}.

In this section, we explore another approach to Type-2 computability based on realizability. The primary motivation for doing so is that we will obtain another perspective on computability (\ie, in addition to the topological and order-theoretic ones) that highlights the connection with \emph{constructive} mathematics. Intuitively, we have a constructive object if we can \emph{realize} the object as a \emph{program}. As another source of motivation, it is also possible to give semantics to programming languages directly using the realizability approach~\citep[\eg, see][]{longley1995thesis}. Hence, we will gain another method of giving semantics in addition to the traditional order-theoretic one.

Under the realizability approach, we will approach Type-2 computability using an abstract machine model, \ie, a \emph{partial combinatory algebra} (PCA) as opposed to a concrete machine model (\ie, a Turing machine). A PCA consists of an underlying set $X$ and a partial application function $\cdot : X \times X \rightharpoonup X$ subject to certain laws that ensure \emph{combinatorial completeness}, \ie, that a PCA can simulate untyped lambda calculus. Hence, we can think of a PCA as an algebraic take on substitution. We obtain ordinary Type-1 computability by instantiating a PCA over the naturals $\N$; the partial application function of a PCA $\cdot: \N \times \N \rightharpoonup \N$ can be defined to simulate the computation of partial recursive functions. By extension, we obtain a Type-2 machine by instantiating a PCA over \emph{Baire space} $\mathbb{B} \eqdef \N \rightarrow \N$; the partial application function of a PCA $\cdot: \mathbb{B} \times \mathbb{B} \rightharpoonup \mathbb{B}$ can be defined to simulate the computation over streams of naturals. In the rest of this section, our goal is to unpack the (well-known) connection between computability and constructive mathematics via realizability, and to show that the base spaces and constructions that are useful for giving semantics to probabilistic programs with continuous distributions can be realized appropriately.

\paragraph{Overview}
The phrase we have in mind is: ``Computability is the realizability interpretation of constructive mathematics"~\citep{bauer2005c2c}. The high-level idea is to encode familiar mathematical objects in an appropriate logic and derive computability as a consequence of having a sound interpretation. Programming up mathematical spaces and their operations will then correspond to encoding the space and their operations in the logic.
\begin{enumerate}[noitemsep,align=left]
    \item[(\emph{Logic})] The logic for our setting is \emph{elementary analysis}~\citep[\eg, see][Sec. 1.3.3]{lietz2004thesis} called $EL$. $EL$ extends an intuitionistic predicate logic with (1) Heyting arithmetic, (2) a sort for Baire space $\BaireTy$ for encoding continuum-sized objects, and (3) primitive-recursion and associated operators.
    \item[(\emph{Semantics})] The semantics for this setting includes the category $\textbf{Asm}(\cK_2)$ of assemblies over Kleene's second algebra $\cK_2$ (\ie, a PCA over Baire space) and the full subcategory $\textbf{Mod}(\cK_2)$ of modest sets over $\cK_2$. For more details on assemblies and modest sets, we refer the reader to the relevant literature~\citep[\eg, see][]{streicher2008realize,bauer2000thesis,birkedal1999thesis}. For our purposes, it suffices to recall that a modest set can be identified with a represented space and that an assembly is a represented space with a \emph{multi-representation}. Hence, modest sets model datatypes and assemblies model intuitionistic logic.
\end{enumerate}

Because we take a constructive vantage point, we will need to check that the semantics induced by the relevant encodings of familiar mathematical objects in the logic coincides with the usual interpretation. For our purposes, this means checking that encodings of objects such as reals and distributions in $EL$ produce the expected semantics. Towards this end, recall that we can associate a theory of effectivity with a space by defining it as a quotient of Baire space $\quot{\baire}{\relmv}$ by a \emph{partial equivalence relation} (PER) $\relmv$. A quotient by a PER allows us to construct quotients and subsets of Baire space in one go. We recall the conditions required of the relation $\relmv$ for the constructive encoding to coincide with the classical interpretation below.
\begin{definition}
  \citep[][Prop. 3.3.2]{lietz2004thesis}\textbf{.}
  We write $\relmv^*$ if
  \begin{enumerate}[noitemsep,align=left]
  \item[(\emph{RF conservative class})] antecedents of implications contained in $\relmv$ are almost negative;\footnote{More formally, whenever $A \rightarrow B$ is a subformula of $\relmv$, then the antecedant $A$ is almost negative. As a reminder, a formula is \emph{almost negative} if it only contains existential quantifiers in front of prime (\ie, atomic) formulas.}
  \item[(\emph{partial equivalence relation})] $EL \vdash \texttt{sym}(\relmv) \synand \texttt{trans}(\relmv)$ where $\texttt{sym}(\relmv) \eqdef \forall \alpha\,\beta: \BaireTy \ldotp \alpha \relmv \beta \syniff \beta \relmv \alpha$ and $\texttt{trans}(\relmv) \eqdef \forall \alpha\,\beta\,\gamma: \BaireTy \ldotp \alpha \relmv \beta \synrarr \beta \relmv \gamma \synrarr \alpha \relmv \gamma$; and
  \item[(\emph{stability})] $EL \vdash \forall x \, y : \BaireTy. \, \lnot \lnot (x \relmv y) \synrarr x \relmv y$.
  \end{enumerate}
  \label{defn:relgood}
\end{definition}
\noindent Now we recall a sufficient condition for the constructive interpretation to coincide with the classical interpretation.
\begin{proposition}
\emph{\citep[][Prop. 3.3.2]{lietz2004thesis}\textbf{.}} If $\relmv^*$, then the interpretations of $\quot{\baire}{\relmv}$ in the categories $\textbf{Asm}(\cK_2)$ and $\textbf{Asm}_t(\cK_2)$ (\ie, the truth or classical interpretation) yield computably equivalent realizability structures.
\end{proposition}
\label{prop:realize:equiv}

\paragraph{Encodings}
Before proceeding to the encodings of the sets of interest in $EL$, we define two enumerations that will be useful for constructing the encodings. Let $\pi_1\langle n, m \rangle = n$ and $\pi_2\langle n, m \rangle = m$ so that they are pairing functions on naturals (e.g., Cantor pairing function). We also overload the notation $\langle \alpha, \beta \rangle$ to pair $\alpha \in \baire$ and $\beta \in \baire$.
\begin{enumerate}[noitemsep,align=left]
    \item[(\emph{Integers})] Encode the integers as
    \[
    \Z = \quot{\N \times \N}{=_\N}
    \]
    where $\langle a, b \rangle =_\N \langle c, d\rangle$ if $a - d = c - b$~\citep[\eg, as in][Sec. 5.5.1]{bauer2000thesis}. In words, we can think of an integer as a difference of two naturals. We write $\IntTy$ to refer to the enumeration on $\N \times \N$.
    \item[(\emph{Rationals})] Encode the rationals as
    \[
    \Q = \quot{\Z \times (\N \backslash \set{0})}{=_\Q}
    \]
    where $\langle p, q \rangle =_\Q \langle s, t \rangle$ if $p \cdot t = s \cdot q$~\citep[\eg, as in][Sec. 5.5.1]{bauer2000thesis}. In words, we can think of a rational as a ratio of an integer and a non-negative natural. We write $\RatTy$ to refer to the enumeration on $\Z \times (\N \backslash \set{0})$. We write $\leq_\Q$ and $<_\Q$ to implement $\leq$ and $<$ respectively on rationals.\footnote{Note that we have that $\langle p, q \rangle < \langle s, t \rangle$ if $p \cdot t < s \cdot q$~\citep[\eg, as in][Sec. 5.5.1]{bauer2000thesis}.}
    \item[(\emph{Non-negative rationals})] Encode the non-negative rationals similarly to the rationals, where we replace $\Z$ with $\N$. We write $\NonNegRatTy$ to refer to the enumeration on $\N \times (\N \backslash \set{0})$. We write $<_{\Q^+}$ to implement $<$ on the non-negative rationals.
\end{enumerate}

We now encode the base spaces as quotients of Baire space. In defining the quotient $\relmv$, it is helpful to recall the encoding of the space first. For example, a \emph{lower real} is an encoding of a real that enumerates all of its rational lower bounds. Hence, two lower reals will be related if their encodings enumerate the same lower bounds. As another example, we can encode reals as a fast Cauchy sequences. Hence, two reals will be related if their fast Cauchy sequences are suitably close to one another. We summarize useful quotient encodings of base spaces below.
\begin{proposition}\hfill
\begin{enumerate}[noitemsep,align=left]
    \item[(\emph{Sierpinski})] Let $\alpha \relmv_\sierp \beta$ if $(\forall n: \NatTy\ldotp \alpha \, n = 0) \syniff (\forall n: \NatTy\ldotp \beta \, n = 0)$.
    \item[(\emph{Lower real})] Let $\alpha \relmv_{\LR} \beta$ if $\forall q: \RatTy\ldotp (\forall n: \NatTy\ldotp q <_\Q \alpha \, n) \syniff (\forall n: \NatTy\ldotp q <_\Q \beta \, n)$.
    \item[(\emph{Lower non-negative real})] Let $\alpha \relmv_{\LR^+} \beta$ if $\forall q: \NonNegRatTy\ldotp (\forall n: \NatTy\ldotp q <_{\Q^+} \alpha \, n) \syniff (\forall n: \NatTy\ldotp q <_{\Q^+} \beta \, n)$.
    \item[(\emph{Upper real})] Let $\alpha \relmv_{\UR} \beta$ if $\forall q: \RatTy\ldotp (\forall n: \NatTy\ldotp \alpha \, n <_\Q q) \syniff (\forall n: \NatTy\ldotp \beta \, n <_\Q q)$.
    \item[(\emph{Lifted partial real})] Let $\langle \alpha_l, \alpha_u, \alpha_<, \alpha_> \rangle \relmv_{\Tilde{\R}} \langle \beta_l, \beta_u, \beta_<, \beta_> \rangle$ if $\alpha_l \relmv_{\R_<} \beta_l \,\land\, \alpha_u \relmv_{\R_>} \beta_u \,\land\, \alpha_< \relmv_{\sierp} \beta_< \,\land\, \alpha_> \relmv_{\sierp} \beta_>$.
    \item[(\emph{Real})] Let $\alpha \relmv_\R \beta$ if $\forall n: \NatTy\ldotp |\alpha \, n - \beta \, n| \leq_\Q 2^{-n + 2}$.
\end{enumerate}
We have $\relmv_\sierp^*$, $\relmv_{\LR}^*$, $\relmv_{\LR^+}^*$, $\relmv_{\UR}^*$, $\relmv_{\tilde{\R}}^*$, and $\relmv_\R^*$.
\end{proposition}
\noindent It is largely straightforward to check that $\relmv^*$ holds for the $\relmv$ defined above.\footnote{For Sierpinski, see~\citet[Defn. 3.2.4]{lietz2004thesis}. For reals, see~\citet[Sec. 5.5.2]{bauer2000realizability}. It is also useful to recall the notion of a \emph{negative formula}~\citep[][pg. 92]{bauer2000realizability} for checking the stability of $\relmv$.} Next, we state that semantic constructs can be encoded as quotients of Baire space as well.
\begin{proposition}
Suppose $\relmv_X^*$ and $\relmv_Y^*$.
\begin{enumerate}[noitemsep,align=left]
    \item[(\emph{Lift})] Let $\langle \alpha_C, \alpha_X \rangle \relmv_\bot \langle \beta_C, \beta_X \rangle$ if $\alpha_C \relmv_{\sierp} \beta_C \land \alpha_X \relmv_X \beta_X$.
    \item[(\emph{Product})] Let $\langle \alpha_X, \alpha_Y \rangle \relmv_{X \times Y} \langle \beta_X, \beta_Y \rangle$ if $\alpha_X \relmv_X \beta_X \land \alpha_Y \relmv_Y \beta_Y$.
    \item[(\emph{Function})] Let $\alpha \relmv_{X \rightarrow Y} \beta$ if $\forall \gamma: \BaireTy, \, \alpha \,|\, \gamma \relmv_Y \beta \,|\, \gamma$ where $\alpha \,|\, \gamma$ applies $\alpha$ to $\gamma$ (in $\cK_2$).
\end{enumerate}
We have $\relmv_\bot^*$, $\relmv_{X \times Y}^*$, and $\relmv_{X \rightarrow Y}^*$.
\end{proposition}
\noindent It is straightforward to check that $\relmv^*$ for the $\relmv$ defined above.

We end by encoding valuations as quotients of Baire space. First, we need an enumeration of the open sets of a topological space. For a topological space $(X, \cO(X))$, we can encode the collection of open sets as the function space $X \rightarrow \sierp$. As the measure of an open set is lower-semi computable (Proposition~\ref{prop:sem:valuations:int}), a valuation can be encoded as an enumeration of pairs of a basic open and a non-negative lower real. For a countably based topological space with basis $\cB(X)$, we have $\cB(X) \cong \N$; hence, we can code a valuation as a sequence of non-negative lower reals.
\begin{proposition}
Let $\langle \alpha_1, \alpha_2, \dots \rangle \relmv_{\cV(X)} \langle \beta_1, \beta_2, \dots \rangle$ if $\forall n: \NatTy \ldotp \alpha_n \relmv_{\R^+_<} \beta_n$. Then $\relmv_{\cV(X)}^*$.
\end{proposition}

\paragraph{Summary}
In summary, one view of what we have just seen is that we can use $EL$ as a ``programming language" (\ie, a constructive logic as opposed to Haskell) for coding up mathematical structures relevant for probabilistic programs that have a notion of effectivity associated with them. In particular, the witnesses in the semantics of $EL$ are given by elements of a PCA and modest sets over $\cK_2$ can be identified with represented spaces~\citep[see][Sec. 8]{battenfeld2007convenient}.

\subsection{Alternative Approaches}
\label{subsec:math:alt}

Probabilistic programs have a long history, and indeed, many structures have been proposed for modeling their semantics. Naturally, the choice of mathematical structure affects the language features that we can model. We close this section by reviewing a few of these alternative approaches as a point of comparison to the perspective given here that emphasizes Type-2 computability. We will focus on denotational approaches. There are also operational approaches to modeling the semantics of probabilistic programs~\citep[\eg, see][]{park2005sampling,lago2012probabilistic}.

One natural idea is to extend semantics based on CPOs to the probabilistic setting by putting distributions on CPOs. \citet{saheb1978plcf} develops a probabilistic version of LCF by considering distributions on CPOs corresponding to base types (\ie, booleans and naturals). Saheb-Djahromi also gives operational semantics as a Markov chain (described as a transition matrix) and shows that the operational semantics is equivalent to the denotational semantics. \citet{jones1989thesis}, in her seminal work, develops the theory of valuations on CPOs to further the study of distributions on CPOs via a \emph{probabilistic powerdomain} $\cP$. The probabilistic powerdomain is not closed under the function space;  consequently, Jones interprets the function space $D \arrdom[] E$ probabilistically as $D \arrdom[] \cP(E)$ (not $\cP(D) \arrdom[] P(E)$).

Instead of taking order-theoretic structure as primary and extending it with probabilistic concepts, another idea is to take the probabilistic structure as primary and derive structure that models programming language constructs (\eg, order-theoretic structure to model recursion). \citet{kozen1981semantics} takes a structure amenable for modeling probability as primary (\ie, Banach spaces) and imposes order-theoretic structure. This approach supports standard continuous distributions, although it does not support higher-order functions. In addition to the distributional semantics, Kozen also gives a sampling semantics and shows it equivalent to the distributional semantics. \citet{danos2011pcs} identify the category of probabilistic coherence spaces (PCSs) and use it to give denotational semantics to a probabilistic variant of PCF extended with (countable) choice. Hence, their approach supports discrete distributions. \citet{ehrhard2014probabilistic} show that PCSs provide a fully abstract model for probabilistic PCF so that the connection between the operational and denotational semantics is tight. \citet{ehrhard2018measurable} identify a Cartesian closed category of measurable cones and stable, measurable maps that is also order complete. They also provide an operational sampling semantics and show an adequacy result to link the denotational with operational semantics. This category can be used to model higher-order probabilistic languages with continuous distributions and recursion. \citet{crubille2018probabilistic} shows that the category of PCSs embeds into the (Cartesian closed) category of measurable cones with stable, measurable maps.

One can also use measure-theoretic structure directly, although the category of measurable spaces with measurable maps is not Cartesian closed so higher-order functions cannot be modeled. \citet{panangaden1999category} identifies a category of stochastic relations and shows how to use it to give denotational semantics to Kozen's first-order while language. The category has measurable spaces as objects and probability kernels as morphisms. Panangaden identifies (partially) additive structure in this category and uses it to interpret fix-points for Kozen's while language. \citet{borgstrom2011measure} also interpret a type as a measurable space and use it to give denotational semantics to a first-order language without recursion based on measure transformers. They also show how to compile this language into a factor graph, which supports inference as well as provides an operational semantics. \citet{staton2017commutative} shows how the category of measurable spaces with $s$-finite kernels can be used to give commutative semantics to a first-order language.

Another interesting approach considers alternatives to a measure-theoretic treatment of probability, but still considers the probabilistic structure as primary. \citet{heunen2017convenient} develop the theory of quasi-Borel spaces, which importantly, form a Cartesian closed category and show how quasi-Borel spaces can be used to model a higher-order probabilistic language with continuous distributions but without recursion. \citet{vakar2019domain} show how to extend quasi-Borel spaces with order-theoretic structure so they can be used to model languages with recursion.
\section{A Semantics for a Core Language}
\label{sec:core}

\emph{What does a semantics for a core language look like?} Our goal in this section is to use the mathematical structures (\ie, topological domains) we reviewed in the previous section to model a PCF-like language extended with reals and continuous distributions (via a probability monad) called $\langprob$. We begin by introducing the syntax and statics of $\langprob$ (Section~\ref{subsec:core:statics}). As we might expect, the language features that we can model are restricted to the structure of the relevant topological domains. For instance, as we only define a probability monad on countably based spaces, the probability monad in $\langprob$ will be restricted to supporting only distributions on countably based spaces. This includes distributions on reals and products of countably based spaces, but does not include function spaces (although the language itself contains higher-order functions). Next, we give both (algorithmic) sampling and distributional semantics to $\langprob$ (Section~\ref{subsec:core:sem}). This illustrates more concretely the connection between the semantics and the library implementation of computable distributions. The structure of the semantics follows the usual one for PCF. Finally, we can use the core language and its semantics to resolve the semantic issues we raised when we sketched a library for computable distributions (Section~\ref{subsec:core:reason}).

\subsection{Syntax and Statics}
\label{subsec:core:statics}

\begin{figure}[t]
    \centering
\begin{align*}
\typ & ::= \natty \bnfsep \typ \arrty \typ \bnfsep \typ \prodty \typ \bnfsep \realty \bnfsep \distty{\typ} \\
M & ::= \texttt{O} \bnfsep \texttt{succ} \bnfsep \texttt{pred} \bnfsep \ifzexp{M}{M}{M} \tag{PCF-1} \\
& \quad \bnfsep x \bnfsep \absexp{x: \typ}{M} \bnfsep M \, N \bnfsep \texttt{fix} \, M \tag{PCF-2} \\
& \quad \bnfsep (M, M) \bnfsep \fstexp{M} \bnfsep \sndexp{M} \tag{products} \\
& \quad \bnfsep \mlhighlight{r} \bnfsep \mlhighlight{\realprimkw} \tag{reals} \\
& \quad \bnfsep \mlhighlight{\distmv} \bnfsep \mlhighlight{\retexp{M}} \bnfsep \mlhighlight{\bindexp{x}{M}{M}} \tag{distributions}
\end{align*}
    \caption{$\langprob$ extends a PCF-like language with products, reals, and distributions using a probability monad. The constructs for reals and distributions are shaded.}
    \label{fig:core:statics:syntax}
\end{figure}

\paragraph{Syntax}
The language $\langprob$ extends a PCF-like language with reals and distributions (Figure~\ref{fig:core:statics:syntax}). The terms on lines \emph{PCF-1} and \emph{PCF-2} are standard PCF terms. The terms on the line marked \emph{products} extend PCF with the usual constructions for pairs; $(M, N)$ forms a pair of terms $M$ and $N$, $\fstexp{M}$ takes the first projection of the pair $M$, and $\sndexp{M}$ takes the second projection of the pair $M$. The terms on the line marked \emph{reals} add syntax for (1) constant reals $r$ and (2) the application of primitive real functions $\realprimkw$. The terms on the line marked \emph{distributions} add syntax for (1) primitive distributions $\distmv$ and (2) return $\retexp{M}$ and bind $\bindexp{x}{M}{N}$ for an appropriate probability monad.

\begin{figure}[t]
\begin{mathpar}
\boxed {\vdash_D \tau} \hfill \text{Well-formed distribution type} \\
\inferrule[]
{\\}
{\vdash_D \natty} \and
\inferrule[]
{\\}
{\vdash_D \realty} \and
\inferrule[]
{\vdash_D \typ_1 \and \vdash_D \typ_2}
{\vdash_D \typ_1 \prodty \typ_2} \and
\end{mathpar}
\begin{mathpar}
\boxed {\tyctx \vdash M : \tau} \hfill \text{Expression typing judgement} \\
\inferrule[]
{\\}
{\tyctx \vdash \texttt{O} : \natty}\and
\inferrule[]
{\\}
{\tyctx \vdash \texttt{succ} : \natty \arrty \natty}\and
\inferrule[]
{\\}
{\tyctx \vdash \texttt{pred} : \natty \arrty \natty}\and
\mlhighlight{\inferrule[]
{\gctx(r) = \realty}
{\tyctx \vdash r : \realty}}\and
\mlhighlight{\inferrule[]
{\gctx(\realprimkw) = \realty^n \rightarrow \realty}
{\tyctx \vdash \realprimkw : \realty^n \rightarrow \realty}}\and
\inferrule[]
{\tyctx \vdash M_1 : \natty \and \tyctx \vdash M_2, M_3 : \typ}
{\tyctx \vdash n : \ifzexp{M_1}{M_2}{M_3} : \typ}\and
\inferrule[]
{\tyctx(x) = \typ}
{\tyctx \vdash x : \typ}\and
\inferrule[]
{\tyctx, x : \typ_1 \vdash M : \typ_2}
{\tyctx \vdash \absexp{x: \typ_1}{M} : \typ_1 \arrty \typ_2}\and
\inferrule[]
{\tyctx \vdash M_1 : \typ_2 \arrty \typ \and \tyctx \vdash M_2 : \typ_2}
{\tyctx \vdash M_1 \; M_2 : \typ}\and
\inferrule[]
{\tyctx \vdash M : \typ \arrty \typ}
{\tyctx \vdash \texttt{fix} \, M : \typ}\and
\inferrule[]
{\tyctx \vdash M_1 : \typ_1 \and \tyctx \vdash M_2 : \typ_2}
{\tyctx \vdash (M_1, M_2) : \typ_1 \prodty \typ_2}\and
\inferrule[]
{\tyctx \vdash M : \typ_1 \prodty \typ_2}
{\tyctx \vdash \fstexp{M} : \typ_1}\and
\inferrule[]
{\tyctx \vdash M : \typ_1 \prodty \typ_2}
{\tyctx \vdash \sndexp{M} : \typ_2}\and
\mlhighlight{\inferrule[]
{\gctx(\distmv) = \distty{\typ} \and \vdash_D \typ}
{\tyctx \vdash \distmv : \distty{\typ}}}\and
\mlhighlight{\inferrule[]
{\tyctx \vdash M : \typ \and \vdash_D \typ}
{\tyctx \vdash \retexp{M} : \distty{\typ}}}\and
\mlhighlight{\inferrule[]
{\tyctx \vdash M_1 : \distty{\typ_1} \and \tyctx, x : \typ_1 \vdash M_2 : \distty{\typ_2} \and \vdash_D \typ_1, \typ_2}
{\tyctx \vdash \bindexp{x}{M_1}{M_2} : \distty{\typ_2}}}\and
\end{mathpar}
\caption{The type-system for $\langprob$. The judgement $\vdash_D \tau$ checks that distribution types are well-formed. The judgement $\tyctx \vdash M : \tau$ checks that expressions are well-typed. The judgement is parameterized by a context $\gctx$ (omitted), which contains that types of primitive distributions and functions. The typing rules for reals and distributions are shaded.}
\label{fig:sem:type}
\end{figure}

\paragraph{Statics}
Like PCF, $\langprob$ is a typed language. In addition to PCF types (\ie, $\natty$ and $\typ \arrty \typ$), $\langprob$ includes the type of products ($\typ \prodty \typ$), reals ($\realty$), and distributions ($\distty{\typ}$). Figure~\ref{fig:sem:type} summarizes the type-system for $\langprob$. The expression typing judgement $\tyctx \vdash M : \typ$ is parameterized by a context $\gctx$ (omitted in the rules) that contains the types of primitive distributions and functions.\footnote{The full expression typing judgement would be written $\gctx; \tyctx \vdash M : \typ$. We omit $\gctx$ because it is constant across typing rules.} The typing rules for the fragments marked \emph{PCF-1}, \emph{PCF-2}, and \emph{products} is standard. The typing rules for the fragments marked \emph{reals} and \emph{distributions} are not surprising; nevertheless, we go over them as the constructs are less standard.

As expected, constant reals $r$ are assigned the type $\realty$. Primitive operations on reals $\realprimkw$ (for real operation) have the type $\realty^n \rightarrow \realty$ where $\realty^n \eqdef \realty \times \dots \times \realty$ ($n$-times).

For expressions that operate on distributions, the judgement $\vdash_D \typ$ additionally enforces that the involved types are well-formed. The distribution type $\distty{\typ}$ is well-formed if the space denoted by $\typ$ is a computable metric space (Definition~\ref{def:crev:real:cms}). This includes the type $\natty$, the type $\realty$ (Example~\ref{ex:crev:real:real}), and products of well-formed types $\typ_1 \prodty \typ_2$.\footnote{As a reminder, we can also support distributions on any countably-based space (\eg, distributions on distributions), but restrict our attention to these types for simplicity.}

Given a term $M$ that has a well-formed type, the construct $\retexp{M}$ corresponds to return in a probability monad and returns a point-mass centered at $M$. The typing rule for $\bindexp{x}{M}{N}$ is the usual one for bind in a probability monad. The rule first checks that $M$ has type $\distty{\typ_1}$ and that $\typ_1$ is well-formed. Next, the rule checks that $N$ under a typing context extended with $x: \typ_1$ has type $\distty{\typ_2}$ and that $\typ_2$ is well-formed. The result is an expression of type $\distty{\typ_2}$.

\subsection{Semantics}
\label{subsec:core:sem}

\begin{figure}[t]
    \centering
\begin{align*}
  \typD{\natty} & \eqdef \liftdom{\N} \\
  \typD{\typ_1 \arrty \typ_2} & \eqdef \liftdom{(\typD{\typ_1} \arrdom[] \typD{\typ_2})} \\
  \typD{\typ_1 \prodty \typ_2} & \eqdef \liftdom{(\typD{\typ_1} \proddom[] \typD{\typ_2})} \\
  \mlhighlight{\typD{\realty}} & \eqdef \mlhighlight{\tilde{\R}_\bot} \\
  \mlhighlight{\typD{\distty{\typ}}} & \eqdef \mlhighlight{\set{(s, \pshfn_{\typD{\typ}}(s)) \ST s \in \sampfunc(\typD{\typ})}}
\end{align*}
    \caption{The interpretation of types $\typD{\cdot}$ denotes types as topological domains. We have shaded the interpretation of reals and distributions. Note that we are using a call-by-name interpretation.}
    \label{fig:core:sem:typd}
\end{figure}

\paragraph{Interpretation of types}
The interpretation of types $\typD{\typ} \in \textbf{TD}$ interprets a type $\typ$ as a topological domain and is defined by induction on types (Figure~\ref{fig:core:sem:typd}). The interpretation of types is similar to what one obtains from a standard CPO call-by-name interpretation. 

For example, the interpretation of $\natty$ lifts the topological domain $\N$. This is similar to the CPO interpretation of naturals as the lifted naturals. The interpretation of functions and products are the usual call-by-name interpretations, the difference being that we use the topological domain counterparts instead. The interpretation of the type of reals $\realty$ is a lifted partial real $\tilde{\R}_\bot$ (recall Example~\ref{ex:math:dom:partialreal}). The interpretation of the type of distributions $\distty{\typ}$ is a pair of a sampler and a distribution such that the sampler realizes the distribution. The (continuous) function $\pshfn_D: \sampfunc(D) \arrdom[] \probfunc(D)$ computes the pushforward\footnote{We have that $\pshfn_D(s) \eqdef U \mapsto \int \indexp{{U}}{\cdot} \, d\mu_s$ where $\mu_s \eqdef \distFair \circ s^{-1}$.} and converts a sampler into its corresponding valuation. The well-formed distribution judgement $\vdash_D \typ$ ensures that the probability monad $\probfunc$ is applied to only the countably based topological domains.

\begin{figure}[t]
\begin{align*}
\expD{\tyctx \vdash x: \typ} & \eqdef \pi_x \\
\expD{\tyctx \vdash \texttt{zero}: \natty} & \eqdef \liftc \circ \catconst \circ \glob(\texttt{zero}) \\
\expD{\tyctx \vdash \texttt{succ}: \natty \arrty \natty} & \eqdef \liftc \circ \catconst \circ \glob(\texttt{succ}) \\
\expD{\tyctx \vdash \texttt{pred}: \natty \arrty \natty} & \eqdef \liftc \circ \catconst \circ \glob(\texttt{pred}) \\
\expD{\tyctx \vdash \absexp{x: \typ_1}{M}: \typ_1 \arrty \typ_2} & \eqdef \liftc \circ \catcurry(\expD{\tyctx, x: \typ_1 \vdash M: \typ_2}) \\
\expD{\tyctx \vdash M_1 \, M_2: \typ_2} & \eqdef \cateval \circ \langle \unliftf(\expD{\tyctx \vdash M_1: \typ_1 \arrty \typ_2}), \expD{\tyctx \vdash M_2: \typ_1} \rangle \\
\expD{\tyctx \vdash \ifzexp{M_1}{M_2}{M_3}: \typ} & \eqdef \operatorname{if0} \circ \langle \expD{\tyctx \vdash M_1: \natty}, \expD{\tyctx \vdash M_2: \typ}, \expD{\tyctx \vdash M_3: \typ} \rangle \\
\expD{\tyctx \vdash \texttt{fix} \, M: \typ} & \eqdef \catfix \circ \unliftf(\expD{\tyctx \vdash M: \typ \arrty \typ}) \\
\expD{\tyctx \vdash (M_1, M_2): \typ_1 \prodty \typ_2} & \eqdef \liftc \circ \langle \expD{\tyctx \vdash M_1: \typ_1}, \expD{\tyctx \vdash M_2: \typ_2}) \rangle \\
\expD{\fstexp{\tyctx \vdash M: \typ_1}} & \eqdef \pi_1 \circ \unliftf \circ \expD{\tyctx \vdash M: \typ_1 \prodty \typ_2} \\
\expD{\sndexp{\tyctx \vdash M: \typ_2}} & \eqdef \pi_2 \circ \unliftf \circ \expD{\tyctx \vdash M: \typ_1 \prodty \typ_2} \\
\mlhighlight{\expD{\tyctx \vdash r: \realty}} & \eqdef \mlhighlight{\liftc \circ \catconst \circ \glob(r)} \\
\mlhighlight{\expD{\tyctx \vdash \realprimkw: \realty^n \rightarrow \realty}} & \eqdef \mlhighlight{\liftc \circ \catconst \circ \glob(\realprimkw)} \\
\mlhighlight{\expD{\tyctx \vdash \distmv: \distty{\typ}}} & \eqdef \mlhighlight{\liftc \circ \catconst \circ \glob(\distmv)} \\
\mlhighlight{\expD{\tyctx \vdash \retexp{M} : \distty{\typ}}} & \eqdef \mlhighlight{\langle \detfn \circ f, \mret \circ f \rangle \mbox{  where $f = \expD{\tyctx \vdash M: \typ}$}} \\
\mlhighlight{\expD{\tyctx \vdash \bindexp{x}{M_1}{M_2} : \distty{\typ_2}}} & \eqdef \mlhighlight{\langle \sampfn \circ \langle \pi_1 \circ f, \pi_1 \circ \catcurry(g) \rangle}, \\ 
  & \phantom{\eqdef} \,\, \mlhighlight{(\pi_2 \circ f) \mbind (\pi_2 \circ g) \rangle} \\
  & \phantom{\eqdef} \quad \mlhighlight{\mbox{where $f = \expD{\tyctx \vdash M_1: \distty{\typ_1}}$}} \\
  & \phantom{\eqdef} \quad \mlhighlight{\mbox{where $g = \expD{\tyctx, x: \typ_1 \vdash M_2: \typ_2}$}}
\end{align*}
\caption{The denotational semantics of $\langprob$ is given by induction on the typing derivation (semantics of additional constructs are shaded). The structure of the semantics is similar to one where we use CPOs. $\glob$ is a global environment used to interpret constants. The function $\pi_x$ projects the variable $x$ from the environment.}
\label{fig:sem:densem}
\end{figure}

\paragraph{Denotation function}
The expression denotation function $\expD{\tyctx \vdash M : \typ} : \typD{\tyctx} \arrdom[] \typD{\typ}$ (see Proposition~\ref{prop:sem:expdwd}) is defined by induction on the typing derivation and is summarized in Figure~\ref{fig:sem:densem}. It is parameterized by a global environment $\glob$ that interprets constant reals $r$, primitive functions $\realprimkw$, and primitive distributions $\distmv$. The global environment $\glob$ should be well-formed (defined shortly) with respect to the global context $\gctx$ used in the expression typing judgement. After we introduce the notion of a well-formed global environment, we walk though the semantics and connect it with the library implementation, with a particular focus on the relation between a sampling and distributional view of probabilistic programs.

\paragraph{Well-formed global environment}
To ensure that we do not introduce non-computable constants into $\langprob$ (\eg, non-computable operations on reals $\realprimkw$) and that the constants have the appropriate types, the global environment $\glob$ should be well-formed with respect to the global context $\gctx$. To distinguish the semantic value obtained from a global environment lookup from the syntax, we will put a bar over the constant (\eg, $\glob(r) = \bar{r}$) to refer to the semantic value. We say that $\glob$ is well-formed with respect to $\gctx$, written $\gctx \vdash \glob$, if the conditions below hold.
\begin{itemize}[noitemsep,align=left]
\item[(\emph{real-wf})] For any $r \in \dom(\gctx)$, $\glob(r)$ is the realizer of a real $\bar{r}$ when $\gctx(r) = \realty$.
\item[(\emph{dist-wf})] For any $\distmv \in \dom(\gctx)$, $\glob(\distmv)$ is the name of a pair $\overline{\distmv}$ that realizes a sampler over values in $\typD{\typ}$ and the corresponding distribution when $\gctx(\distmv) = \distty{\typ}$.
\item[(\emph{rop-wf})] For any $\realprimkw \in \dom(\gctx)$, the corresponding semantic function $\overline{\realprimkw}$ is strict, continuous on its domain, and a $n$-ary real-valued function on reals when $\gctx(\realprimkw) = \realty^n \rightarrow \realty$.
\end{itemize}

\paragraph{Denotation function and sampling}
The denotation of terms corresponding to the PCF fragment are standard. Hence, we will focus on the constructs $\langprob$ introduces. The denotation of a constant real $r$ is a global environment lookup.
\[
\expD{\tyctx \vdash r: \realty} \eqdef \liftc \circ \catconst \circ \glob(r)
\]
By the well-formedness of the global environment, $\glob(r)$ will have a realizer. Likewise, the denotation of a primitive function on reals $\realprimkw$ is a global environment lookup and corresponds to a representation of the code implementing the function.
\[
\expD{\tyctx \vdash \realprimkw: \realty^n \arrty \realty} \eqdef \liftc \circ \catconst \circ \glob(\realprimkw)
\]
The well-formedness of the global environment $\glob$ enforces these conditions. Our next task is to explain the denotation of distribution constructs in $\langprob$.

As a reminder, the interpretation of types is a pair of a sampler and the distribution that it realizes. As we will see shortly, the semantics of the sampling component and the semantics of the distribution component do not depend on one another (besides the fact that we want the distribution to be realized by the sampler). Hence, we could have given two different semantics and related them. Nevertheless, in this form, we will obtain that the
valuation is the pushforward along the sampler, and consequently, make the connection between what is given by a distributional semantics and what was implemented in the sampling library. We walk through the distribution constructs now.

The denotation of a constant primitive distribution $\distmv$ is a global environment lookup. Note that the interpretation of $\distty{\typ}$ is a pair of a sampler and valuation so the lookup should also produce a pair.
\[
\expD{\tyctx \vdash \distmv: \distty{\typ}} \eqdef \liftc \circ \catconst \circ \glob(\distmv) \\
\]
The denotation of $\retexp{M}$ produces a pair of a sampler that ignores the input bit-randomness and a point mass valuation centered at $M$.
\[
\expD{\tyctx \vdash \retexp{M} : \distty{\typ}} \eqdef \langle \detfn \circ f, \mret \circ f \rangle \mbox{  where $f = \expD{\tyctx \vdash M: \typ}$}
\]
The meaning of $\bindexp{x}{M}{N}$ also gives a sampler and a valuation.
\begin{multline*}
\expD{\tyctx \vdash \bindexp{x}{M_1}{M_2} : \distty{\typ_2}} \eqdef \\ \langle \sampfn \circ \langle \pi_1 \circ f, \pi_1 \circ \catcurry(g) \rangle, (\pi_2 \circ f) \mbind (\pi_2 \circ g) \rangle
\end{multline*}
where $f = \expD{\tyctx \vdash M_1: \distty{\typ_1}}$ and $g = \expD{\tyctx, x: \typ_1 \vdash M_2: \typ_2}$.
Under the sampling view, we use $\sampfn$ to compose the sampler obtained by $\pi_1 \circ f$ with the function $\pi_1 \circ g$. Under the valuation component, we reweigh $\pi_2 \circ g$ according to the valuation $\pi_2 \circ f$ using monad bind $\mbind$ from $\probfunc$. We can check that the valuation given by the semantics is indeed the pushforward along the sampler.
\begin{proposition}[Push] 
  Let $D$ and $E$ be countably based topological predomains ($\text{qcb}_0$ spaces more generally).
\begin{enumerate}[noitemsep]
\item $\pshfn_D(\detfn(d)) = \mret(d)$ for any $d \in D$.
\item $\pshfn_E(\sampfn(s)(f)) = \pshfn_D(s) \mbind \absD{v}{\pshfn_E(f(v))}$ for any $s \in \sampfunc(D)$ and $f : D \arrdom[] \sampfunc(E)$.
\end{enumerate}
\label{lem:sem:abspush}
\end{proposition}
\noindent In the case of bind (the second item), it is necessary that the split operation used in $\sampfn$ (Section~\ref{subsec:math:sampling}) produces an independent stream of bits.

We end by checking that the expression denotation function is well-defined.
\begin{proposition}[Well-defined]
  The expression denotation function $\expD{\cdot}$ is well-defined, \ie, $\expD{\tyctx \vdash M : \typ} : \typD{\tyctx} \arrdom[] \typD{\typ}$ for any well-typed term $\tyctx \vdash M : \typ$ and well-formed global environment $\gctx \vdash \glob$.
  \label{prop:sem:expdwd}
\end{proposition}
\noindent The structure of the argument showing that the expression denotation function is well-defined is similar to the argument for showing that the CPO semantics of PCF is well-defined. The interesting cases correspond to $\retexp{M}$ and $\bindexp{x}{M_1}{M_2}$ where we need to relate the sampling component with the valuation it denotes, which is given by Proposition~\ref{lem:sem:abspush}.

\subsection{Reasoning About Programs}
\label{subsec:core:reason}

We now return to resolving some semantic issues that were raised when we used the library to implement distributions. Throughout this section, we overload $\expD{\tyctx \vdash M: \distty{\typ}}$ to mean $\pi_2 \circ \expD{\tyctx \vdash M: \distty{\typ}}$ so that it just provides the distributional view. As shorthand, we write $\expDe{\cdot}$ instead of $\expD{\cdot}(\env)$ where the meta-variable $\env$ ranges over environments.

\paragraph{Reasoning about distributions}
We first show that the encoding of the standard geometric distribution is correct. Let $\mu_B$ be an unbiased Bernoulli distribution and $\mu^n$ correspond to $n$ un-foldings of $\texttt{stdGeometric}$:
\begin{align*}
  \expDe{\texttt{stdGeometric}}(U) & = \sup_{n \in \N} \int \left(v \mapsto \begin{cases}
    \indexp{U}{1} & \mbox{$v = t$} \\
    \int \absD{w}{\indexp{U}{w + 1}} d\mu^n & \mbox{$v = f$}
    \end{cases} \right) d\mu_B \\
    & = \sup_{n \in \N} (\indexp{U}{1} \frac{1}{2} + \sum_{w = 0}^\infty \indexp{U}{w + 1} \mu^n(\set{w})) \\
    & = \indexp{U}{1} \frac{1}{2} + \sum_{w = 0}^\infty \indexp{U}{w + 1} (\sup_{n \in \N} \mu^n(\set{w})) \,.
\end{align*}
By induction on $n$, we can show that $\mu^n$ is the measure
\[
\mu^n = \set{0} \mapsto 0, \set{1} \mapsto (1/2), \dots, \set{n}
\mapsto (1/2)^n \;.
\]
Hence, we can conclude that $\sup_{n \in \N} \mu^n$ is a geometric distribution and that the encoding of $\texttt{stdGeometric}$ is correct (for any environment $\env$).

\paragraph{Primitive functions}
In our encoding of the standard normal distribution via the Marsaglia polar transformation, we used $\texttt{<}$ as if it had a return type of $\boolty$ even though equality on reals is not computable. Indeed, the well-formedness conditions imposed on the global environment would disallow $\texttt{<}$ at the current type. To resolve the semantics of $\texttt{<}$, we can think in terms of an implementation. In particular, we can encode $\texttt{<}$ as dovetailing computations that semi-decides $x < y$ (\ie, return $\texttt{()}$ if $x < y$ and diverge otherwise) and semi-decides $x > y$ (\ie, returns $\texttt{()}$ if $x > y$ and diverge otherwise)). On the case of equality, which occurs with probability $0$ in the Marsaglia polar transform, the function diverges.

\paragraph{Partiality and divergence}
We investigate the semantics of divergence more closely now. For convenience, we repeat the two expressions from Section~\ref{sec:lib} that provided two differing notions of divergence below.
\begin{hlisting}
botSamp :: (CMetrizable a) => Samp (Approx a)
botSamp = botSamp

botSampBot :: (CMetrizable a) => Samp (Approx a)
botSampBot = mkSamp (\_ -> bot)
    where bot = bot
\end{hlisting}
In the former, we obtain the bottom valuation, which assigns $0$ mass to every open set. This corresponds to the sampling function $\absD{u \in \two^\omega}{\bot}$ and can be interpreted as failing to provide a sampler. In the latter, we obtain the valuation that assigns $0$ mass to every open set, except for the set $\set{\lift{X} \cup \bot}$ which is assigned mass $1$. This corresponds to the sampling function $\absD{u \in \two^\omega}{\lift{\bot}}$ and can be interpreted as providing a sampling function that fails to produce a sample.

As before, we can check that laziness works in the appropriate manner by selectively ignoring the results of draws from the distributions above.

\begin{minipage}{0.45\textwidth}
\begin{hlisting}
alwaysDiv :: Samp Real
alwaysDiv = do
  _ <- botSamp ::
         Samp Real
  stdUniform
\end{hlisting}
\end{minipage}
\begin{minipage}{0.45\textwidth}
\begin{hlisting}
neverDiv :: Samp Real
neverDiv = do
  _ <- botSampBot ::
         Samp Real
  stdUniform
\end{hlisting}
\end{minipage}

\noindent We can check that the denotation of the former is equivalent to that
of $\texttt{botSamp}$: 
\begin{align*}
\expDe{\texttt{alwaysDiv}}(U) & = \int \left( \absD{\cdot}{\mu_{\cU}(U)} \right) \, d\expDe{\texttt{botSamp}} \\
& = 0
\end{align*}
where $\mu_{\cU}$ is the standard uniform distribution. Note that $\expDe{\texttt{botSamp}}$ maps every open set to $0$ so the integral is $0$ as well. However, the denotation of the latter is equivalent to that of $\texttt{stdUniform}$:
\begin{align*}
  \expDe{\texttt{neverDiv}}(U) & = \int \left( \absD{\cdot}{\mu_{\cU}(U)} \right) \, d\expDe{\texttt{botSampBot}} \\
  & = \sup_{s \text{ simple}} \set{\int s \, d\expDe{\texttt{botSampBot}} \ST s \leq \absD{\cdot}{\mu_{\cU}(U)}} \\
  & = \mu_{\cU}(U) \,.
\end{align*}
As a reminder, $\expDe{\texttt{botSampBot}}(U) = 1$ when $U = \typD{\realty}$. Hence, the integral takes its largest value on the simple function\footnote{As a reminder, a simple function in our context is a linear combination of indicator functions on open sets.} $\mu_{\cU}(U) \, \indexp{{\typD{\realty}}}{\cdot}$.

As a final example, consider the program below that uses a coin flip to determine its diverging behavior.
\begin{hlisting}
maybeBot :: Samp Bool
maybeBot = do
  b <- stdBernoulli
  if b then return bot else stdBernoulli
\end{hlisting}
Intuitively, this distribution returns a \emph{sampler that always generates diverging samples} with probability $1/2$ and returns an unbiased Bernoulli distribution with probability $1/2$. If we changed $\texttt{return bot}$ to $\texttt{botSamp}$ as below
\begin{hlisting}
maybeBot' :: Samp Bool
maybeBot' = do
  b <- stdBernoulli
  if b then bot else stdBernoulli
\end{hlisting}
then the semantics would change to a distribution that returns a \emph{diverging sampler} with probability $1/2$ and an unbiased Bernoulli distribution with probability $1/2$.

\paragraph{Independence and commutativity}
In Section~\ref{subsec:lib:ex}, we saw that we could not argue that two distributions that commuted the order in which we sampled independent normal distributions were equivalent. As a reminder, the issue was that commuting the order of sampling meant that the underlying random bit-stream was consumed in a different order. Consequently, the values produced by the two terms may be different. However, as the semantics we just saw relates the sampling view with the distributional view by construction, we can easily see that these two terms will be distributionally equivalent by Fubini.

\section{Bayesian Inference}
\label{sec:infer}

\emph{What are the implications of taking a computable viewpoint for Bayesian inference?} In this section, we discuss the implications of taking a computable viewpoint for Bayesian inference. Perhaps surprisingly, one can show that conditioning is not computable in general. Nevertheless, conditioning in practical settings does not run into these pathologies. It will be important for probabilistic programming languages to support conditioning in these cases. Note that these results say nothing about the efficiency of inference. In practice, we will still need approximate inference algorithms to compute conditional distributions.

\subsection{Conditioning is not Computable}
\label{subsec:infer:notcomp}

\begin{figure}
    \centering
\begin{hlisting}
nonComp :: Samp (Nat, Real)
nonComp = do
  n <- geometric (1/2)
  c <- bernoulli (1/3)
  u <- uniform 0 1
  v <- uniform 0 1
  x <- return (mkApprox
               (\k -> select u v c k (tmHaltsIn n)))
  return (n, x)
    where select u v c k m
            | m > k = nthApprox v k
            | m == k = if c then 1 else 0
            | m < k = nthApprox u (k - m - 1)  
\end{hlisting}
    \caption{A Haskell encoding of a counter-example given by~\citet{ackerman2011noncomputable} that shows that conditioning is not always computable. The function \texttt{tmHaltsIn} in the code outputs the number of steps the $n$-th Turing machine halts in or $\infty$ (for diverges) if the $n$-th Turing machine does not halt. The idea is that an algorithm that could compute this conditional distribution would imply a decision procedure for the Halting problem.}
    \label{fig:infer:notcomp:counterex}
\end{figure}

Figure~\ref{fig:infer:notcomp:counterex} gives an encoding in $\langprob$ of an example by~\citet{ackerman2011noncomputable} that shows that conditioning is not always computable. Similar to other results in computability theory, the example demonstrates that an algorithm computing the conditional distribution would also solve the Halting problem. The function \texttt{tmHaltsIn} accepts a natural $n$ specifying the $n$-th Turing machine and outputs the number of steps the $n$-th Turing machine halts in or $\infty$ (for diverges) if the $n$-the Turing machine does not halt. Upon inspection, we see the function \texttt{nthApprox} produces the binary expansion (as a dyadic rational) of a real, using \texttt{tmHaltsIn} to select different bits of the binary expansion of $u$ or $v$, or the bit $c$ depending on whether the $n$-th Turing machine halts within $k$ steps or not. 

Consider computing the conditional distribution $\PP(\mathbf{N} \given \bX)$, where the random variable $\mathbf{N}$ corresponds to the program variable $n$ and $\bX$ to $x$. Thus, computing the conditional distribution $\PP(\mathbf{N} \given \bX)$ corresponds to determining the value of the program variable $n$ given the value of the program variable $x$. We informally discuss why this distribution is not computable now. Observe that (1) the value of $x$ depends on the value of $n$---whether the $n$-th Turing machine halts within $k$ steps or not for every $k$ (hence whether the $n$-th Turing machine halts or not)---and (2) the geometric distribution is supported on $\N \backslash \set{0}$ so we need to consider every Turing machine. Consequently, we would require a decision procedure for the Halting problem in order to compute the posterior distribution on $n$. Thus, $\texttt{nonComp}$ encodes a computable distribution $\PP(\mathbf{N}, \bX)$ whose conditional $\PP(\mathbf{N} \given \bX)$ is not computable. We refer the reader to the full proof~\citep[see][]{ackerman2011noncomputable} for more details.

\subsection{Conditioning is Computable}
\label{subsec:infer:practice}

\begin{figure}[t]
\begin{hlisting}
module CondLib (BndDens, obsDens) where
import ApproxLib
import CompDistLib  
import RealLib      

newtype BndDens a b =
    BndDens { getBndDens :: (Approx a -> Approx b -> Real, Rat) }

-- Requires bounded and computable density
obsDens :: forall u v y. 
    (CMetrizable u, CMetrizable v, CMetrizable y) =>
    Samp (Approx (u, v)) -> BndDens u y -> Approx y -> Samp (Approx (u, v))
    
-- Extend with more conditioning operators below ...
\end{hlisting}
\caption{An interface for conditioning (module \texttt{CondLib}). The function \texttt{obsDens} enables conditioning on continuous-valued data when a bounded and computable conditional density is available.}
\label{fig:sem:cond}
\end{figure}

Now, we add conditioning as a library to $\langprob$
(Figure~\ref{fig:sem:cond}). $\langprob$ provides only a restricted
conditioning operation $\obsdenslib$, which requires a conditional
density. We will see that the computability of $\obsdenslib$ corresponds
to an effective version of Bayes' rule. We have given only one
conditioning primitive here, but it is possible to identify other
situations where conditioning is computable and add those to the
conditioning library. For example, conditioning on positive
probability events is computable~\citep[see][Prop. 3.1.2]{galatolo2010effective}.

The library provides the conditioning operation $\obsdenslib$, which
enables us to condition on \emph{continuous}-valued data when a bounded
and computable conditional density is available.
\begin{proposition}
\emph{\citep[][Cor. 8.8]{ackerman2011noncomputable}}\textbf{.} Let $\bm{U}$ be a $U$-valued random variable, $\bm{V}$ be a $V$-valued random variable, and $\bm{Y}$ be $Y$-valued random variable, where $\bm{Y}$ is independent of $\bm{V}$ given $\bm{U}$. Let $\bm{U}$, $\bm{V}$, and $\bm{Y}$ be computable. Moreover, let $p_{\bm{Y} \given \bm{U}}(y \given u)$ be a conditional density of $\bm{Y}$ given $\bm{U}$ that is bounded and computable. Then the conditional distribution $\PP[(\bm{U}, \bm{V}) \given \bm{Y}]$ is computable.
\end{proposition}
The bounded and computable conditional density enables the following integral to be computed, which is in essence Bayes' rule. A version of the conditional distribution $\PP((\bm{U}, \bm{V}) \given \bm{Y})$ is
\[
\kappa_{(\bm{U}, \bm{V}) \given \bm{Y}}(y, B) = \frac{\int_B p_{\bm{Y} \given \bm{U}}(y \given u) \, d\mu_{(\bm{U}, \bm{V})}}{\int p_{\bm{Y} \given \bm{U}}(y \given u) \, d\mu_{(\bm{U}, \bm{V})}}
\]
where $B$ is a Borel set in the space associated with $U \times V$ and $\mu_{(\bm{U}, \bm{V})}$ is the joint distribution of $\bm{U}$ and $\bm{V}$.\footnote{As a reminder, $p_{\bm{Y} \given (\bm{U}, \bm{V})}(y \given u, v) = p_{\bm{Y} \given \bm{U}}(y \given u)$ due to the conditional independence of $\bm{Y}$ and $\bm{V}$ given $\bm{U}$. Hence, the conditional density $p_{\bm{Y} \given \bm{U}}(y \given u)$ in the integral written more precisely is $(u, v) \mapsto p_{\bm{Y} \given \bm{U}}(y \given u)$.}

Another interpretation of the restricted situation is that our observations have been corrupted by independent smooth noise~\citep[][Cor. 8.9]{ackerman2011noncomputable}. To see this, consider the following generative model:
\begin{align*}
    (\bm{U}, \bm{V}) & \sim \mu_{(\bm{U}, \bm{V})} \\
    \bm{N} & \sim \mu_{\text{noise}} \\
    \bm{Y} & = \bm{U} + \bm{N}
\end{align*}
where $\mu_{\text{noise}}$ has density $p_{\bm{N}}(\cdot)$. The random variable $\bm{U}$ can be interpreted as the ideal model of how the data was generated and the random variable $\bm{V}$ can be interpreted as the model parameters. The random variable $\bm{Y}$ can then be interpreted as the data we observe that is smoothed by the noise $\bm{N}$ so that $p_{\bm{Y} \given \bm{U}}(y \given u) = p_{\bm{N}}(y - u)$. Notice that the model $(\bm{U}, \bm{V})$ is not required to have a density and can be an arbitrary computable distribution. The idea is that we condition on $\bm{Y}$ (\ie, the smoothed data) as opposed to $\bm{U}$ (\ie, the ideal data) when we compute the posterior distribution for the model parameters $\bm{V}$.\footnote{As an illustrative example, consider the following situation where we hope to model how a scene in an image is constructed so we can identify objects in the image~\citep[see][for a probabilistic programming language designed for scene perception]{kulkarni2015picture}. The model parameters $\bm{V}$ contain all the information describing the objects in the scene, including their optical properties and their positions. The resulting image $\bm{U}$ is then rendered by a graphics engine. The caveat is that the graphics engine uses an enumeration of the Halting set to add artifacts to the image so it is pathological. (Thus, this distribution is a version of $\texttt{nonComp}$.) Instead of attempting to compute the posterior distribution $\PP(\bm{U} \ST \bm{V})$, which is not computable, we smooth out the rendered image $\bm{U}$ with some noise given by $p_{\bm{Y} \given \bm{U}}(y \given u)$. In other words, we apply some filtering to the image $\bm{U}$ so we obtain an image $\bm{Y}$ free of artifacts introduced by the pathological graphics engine. The posterior $\PP((\bm{U}, \bm{V}) \ST \bm{V})$ is then computable. In this example, the posterior would give the positions and optical properties of objects given an image so it could be used in computer vision applications.} Indeed, probabilistic programming systems proposed by the machine learning community impose a similar restriction~\citep[\eg, see][]{goodman2008church,wood2014new}.

Now, we describe $\obsdenslib$, starting with its type signature. Let the type $\bnddensty{\typ}{\typi}$ represent a bounded computable density:
\begin{hlisting}
newtype BndDens a b =
    BndDens { getBndDens :: (Approx a -> Approx b -> Real, Rat) }
\end{hlisting}
Conditioning thus takes a samplable distribution, a bounded computable density describing how observations have been corrupted, and returns a samplable distribution representing the conditional. In the context of Bayesian inference, it does not make sense to condition distributions such as $\texttt{maybeBot}$ that diverge with positive probability. Hence, we do not give semantics to conditioning on those distributions.

The implementation of $\obsdenslib$ is in essence a $\langprob$ program that implements the proof that conditioning is computable in this restricted setting. This is possible because results in computability theory have computable realizers.\footnote{That is, we implement the Type-2 machine code as a Haskell program.}
\begin{hlisting}            
obsDens :: forall u v y. 
    (CMetrizable u, CMetrizable v, CMetrizable y) =>
    Samp (Approx (u, v)) -> BndDens u y -> Approx y -> Samp (Approx (u, v))
obsDens dist (BndDens (dens, bnd)) d =
    let f :: Approx (u, v) -> Real = \x -> dens (approxFst x) d
        mu :: Prob (u, v) = sampToComp dist
        nu :: Prob (u, v) = \bs ->
              let num  = integrateBndDom mu f bnd bs
                  denom = integrateBnd mu f bnd 
              in map fst (cauchyToLU (num / denom))
    in
      compToSamp nu              
\end{hlisting}
The parameter \texttt{dist} corresponds to the joint distribution of the model (both model parameters and likelihood), \texttt{dens} corresponds to a bounded conditional density describing how observation of data has been corrupted by independent noise, and \texttt{d} is the observed data. Next, we informally describe the undefined functions in the sketch. The function \texttt{approxFst} projects out the first component of a product of approximations. The functions \texttt{sampToComp} and \texttt{compToSamp} witness the computable isomorphism between samplable and computable distributions.\footnote{The computable isomorphism relies on the distributions being full-measure. The algorithm is undefined otherwise.} The functions \texttt{integrateBndDom} and \texttt{integrateBnd} compute an integral~\citep[see][Prop. 4.3.1]{hoyrup2009computability}, and correspond to an effective Lebesgue integral. \texttt{cauchyToLU} converts a Cauchy description of a computable real into an enumeration of lower and upper bounds.

Because $\obsdenslib$ works with conditional densities, we do not need to worry about the Borel paradox. The Borel paradox shows that we can obtain different conditional distributions when conditioning on probability zero events~\citep[\eg, see][]{rao2006probability}. To illustrate this, suppose that $\bX$ and $\bY$ are two independent random variables with standard normal distributions. We can ask a (classic) question: ``What is the conditional distribution of $\bY$ given that $\bX = \bY$?"

In statistics, the appropriate response is to notice that the question as posed is ill-formed---one cannot condition on a measure zero event. The well-posed formulation is to define an auxiliary random variable $\mathbf{Z}$ and condition on a constant. For instance, $\mathbf{Z} = \bX - \bY$ conditioned on $\mathbf{Z} = 0$, $\mathbf{Z} = \bY / \bX$ conditioned on $\mathbf{Z} = 1$ , and $\mathbf{Z} = \mathbb{I}_{Y = X}$ conditioned on $\mathbf{Z} = 1$. Remarkably, all three versions lead to different answers~\citep{proschan1998expect}.

A probabilistic programming language that does not provide a notion of random variable such as $\langprob$ will need an alternative method of addressing this issue. Type-2 computability provides a straight-forward answer---it is not possible to create a boolean value that distinguishes two probability zero events in $\langprob$. For instance, the operator $\texttt{==}$ implementing equality on reals returns false if two reals are provably not-equal and diverges otherwise because equality is not decidable.

\section{Summary and Further Directions}
\label{sec:concl}

We hope to have shown that we do not need to sacrifice traditional notions of computation when modeling reals and continuous distributions by keeping their \emph{representations} in mind. The simple observation is that we can ``program" them in a general-purpose programming language. With this in mind, we can now ask a basic question: ``What does it mean for a probabilistic programming language to be Turing-complete?" From the perspective of Type-2 computability, one answer is that such a language can express all \emph{Type-2 computable distributions}, analogous to how a Turing-complete language can express all computable functions. Indeed, this resolution is somewhat tautological!

This answer raises another interesting question related to full-abstraction and universality\footnote{A programming language is universal if all computable elements in the domain of interpretation are definable.} of probabilistic programs. In the standard setting of PCF, one approach to the full-abstraction problem is to add \emph{parallel or} \texttt{por} to the language so that the operational behavior coincides with the denotational semantics. Additionally adding a searching operator \texttt{exists} means that all computable functions will be definable. One may wonder, if an analogous result holds for probabilistic programs. In particular, a universality result would crystallize the thought that Turing-complete probabilistic programming languages express Type-2 computable distributions.

As we are now back on familiar grounds with regards to computability, we can turn our attention to the \emph{design} of probabilistic programming languages. The design of such languages will demand more from a semantics of probabilistic programs. For example, for the purposes of automating Bayesian inference, it is crucial that the inference procedure be \emph{efficient} (and not simply computable). One direction is to find compilation strategies that can efficiently realize Type-2 computable distributions or approximate them (for some notion of approximation) using floating point numbers. Another direction is to consider alternative language designs (in addition to PCF with a probability monad) and the corresponding structures that we will need to model these languages.

\paragraph{Acknowledgments}
We thank our anonymous reviewers for their helpful comments and feedback.


\renewcommand{\refname}{Bibliography}
\bookreferences 
\bibliography{references} \label{refs}
\bibliographystyle{cambridgeauthordate.bst} 
\end{document}